\documentclass[aps,prd,superscriptaddress,showpacs,tighten,nofootinbib]{revtex4}
\usepackage[a4paper, total={6.0in, 9in}]{geometry}
\usepackage{amsfonts,amssymb,latexsym,amsmath} 
\usepackage{bm}
\usepackage{hhline}
\usepackage{longtable}
\usepackage{array} 
\usepackage{color}
\usepackage{ifthen}
\usepackage{epsfig}
 \usepackage{multirow}
\usepackage{enumerate}
\usepackage{dsfont}
\usepackage{psfrag}
\usepackage{float}
\usepackage{epstopdf}
\usepackage{mathtools,slashed}
\usepackage{xcolor,booktabs}

\newcommand{\Tr}{\rm{Tr}}

\newcommand{\be}{\begin{equation}}
 \newcommand{\ee}{\end{equation}}
\newcommand{\ba}{\begin{array}{c}}
 \newcommand{\ea}{\end{array}}
\newcommand{\bea}{\begin{eqnarray}}
 \newcommand{\eea}{\end{eqnarray}}

\begin{document}
\bibliographystyle{unsrt}
\title{Magnetic moments of the spin-$1\over 2$ singly charmed baryons in chiral perturbation theory}

\author{Guang-Juan Wang}\email{wgj@pku.edu.cn}
\affiliation{School of Physics and State Key Laboratory of Nuclear
Physics and Technology, Peking University, Beijing 100871, China}

\author{Lu Meng}\email{lmeng@pku.edu.cn}
\affiliation{School of Physics and State Key Laboratory of Nuclear
Physics and Technology, Peking University, Beijing 100871, China}

\author{Hao-Song Li}\email{haosongli@pku.edu.cn}\affiliation{School of Physics and State Key Laboratory of Nuclear Physics and Technology, Peking University, Beijing 100871, China}

\author{Zhan-Wei Liu}\email{liuzhanwei@lzu.edu.cn}
\affiliation{School of Physical Science and Technology, Lanzhou
University, Lanzhou 730000, China}

\author{Shi-Lin Zhu}\email{zhusl@pku.edu.cn}\affiliation{School of Physics and State Key Laboratory of Nuclear Physics and Technology, Peking University, Beijing 100871, China}\affiliation{Collaborative Innovation Center of Quantum Matter, Beijing 100871, China}

\begin{abstract}

We systematically derive the analytical expressions of the
magnetic moments of the spin-$1\over 2$ singly charmed baryons to the
next-to-next-to-leading order in the heavy baryon chiral
perturbation theory (HBChPT). We discuss the analytical relations between the magnetic moments. We estimate the-low energy constants (LECs) in two scenarios. In the first scenario, we use the quark model and Lattice QCD
simulation results as input. In the second scenario, the heavy quark symmetry is adopted to reduce the number of the independent LECs, which are then fitted using the data from the Lattice QCD simulations. We give the numerical results to the next-to-leading order for the antitriplet
charmed baryons and to the next-to-next-to-leading order for the
sextet states.

\end{abstract}


\maketitle

\section{Introduction}

In the past decades, many heavy baryons and their excitations have
been observed in experiments~\cite{Patrignani:2016xqp}. For instance, the $\Omega_c^*$ was observed in the decay channel $\Omega_c^*\rightarrow \Omega_c \gamma $ by the BABAR Collaboration and confirmed by the Belle Collaboration~\cite{Aubert:2006je,Solovieva:2008fw}. The other two radiative decay processes, $\Xi_c^{'+} \rightarrow \Xi_c^{+} \gamma$  and $\Xi_c^{'0} \rightarrow \Xi_c^0 \gamma$ were also observed in experiments~\cite{Jessop:1998wt,Aubert:2006rv,Yelton:2016fqw}. The electromagnetic processes become more important when strong decay processes are forbidden due to the phase space. These processes provide a platform to study their electromagnetic properties, which are very important to explore the inner structures of the heavy baryons. More experiment data about the magnetic moment and other electromagnetic properties from PANDA, LHCb, BESIII, Belle II and so on are expected
in the future.

The electromagnetic properties of the heavy baryons have attracted the attention of many theorists. Many theoretical models have been adopted to study the electromagnetic properties of the heavy baryons.
The radiative decays of the heavy baryons were investigated with the
chiral symmetry and heavy quark symmetry in Ref.~\cite{Cheng:1992xi}. In Ref.~\cite{Savage:1994wa}, the radiative
decay $\Sigma^{*0}_c\rightarrow \Lambda_c^{0}+\gamma$ was studied in
chiral perturbation theory. In Refs.~\cite{Jiang:2015xqa,Banuls:1999br,Tiburzi:2004mv}, the electromagnetic decays of
the heavy baryons have been calculated in the framework of the heavy
baryon chiral perturbation theory. In Refs.~\cite{Bahtiyar:2015sga,Can:2015exa,Bahtiyar:2016dom,Bahtiyar:2018vub}, the authors studied the
radiative decay and magnetic moments of the heavy baryons with the Lattice QCD simulation. The
radiative decays and magnetic moments of the heavy baryons were also studied using the
heavy quark symmetry~\cite{Tawfiq:1999cf}, various quark models
\cite{Ivanov:1996fj, Ivanov:1999bk, JuliaDiaz:2004vh,
Faessler:2006ft, Albertus:2006ya, Sharma:2010vv, Barik:1984tq,
Wang:2017kfr}, QCD sum rule formalism~\cite{
Agamaliev:2016fou,Aliev:2011bm} and the bag model
\cite{Bernotas:2013eia}. In addition, the magnetic moments of the heavy baryons have been
calculated in the bag model~\cite{Bernotas:2012nz,Bose:1980vy,Simonis:2018rld}, the QCD sum
rules~\cite{ Aliev:2008ay, Aliev:2008sk, Zhu:1997as}, the effective
quark mass and screened charge scheme~\cite{Kumar:2005ei}, the hyper
central model~\cite{Patel:2007gx}, the quark-diquark Model
\cite{Majethiya:2011ry}, the skyrme model~\cite{Oh:1995eu,Oh:1991ws}, the
mean-field approach~\cite{Yang:2018uoj}, the bound-state approach~\cite{Scholl:2003ip}, and the heavy hadron chiral
perturbation theory~\cite{Tiburzi:2004mv,Scimemi:1999td,Banuls:1999mu}. In Ref.
\cite{Can:2013tna}, the magnetic moments and charge radii of the
$\Sigma_c$ and $\Omega_c$ were calculated with the Lattice QCD
simulations.

The chiral perturbation theory (ChPT) is a very useful tool to study
the hadron properties at the low-energy regime. However, the heavy
baryon mass introduces a new large scale and does not vanish in the
chiral limit, which destroys the chiral power counting. Three methods were proposed to overcome this obstacle, the heavy
baryon chiral perturbation theory (HBChPT)~\cite{Jenkins:1990jv,Bernard:1992qa,Hemmert:1997ye}, the infrared regularization (IR) \cite{Becher:1999he}, and the extended-on-mass-shell regularization (EOMS)~\cite{Gegelia:1999gf,Fuchs:2003qc}. The IR and the EOMS are the relativistic formalization of the chiral perturbation theory. They have been used to study the electromagnetic properties of baryons \cite{Geng:2008mf,Xiao:2018rvd,Kubis:2000aa}.  In the HBChPT, the heavy baryon field is decomposed into the ``light" and ``heavy" components. The ``heavy" component is integrated
out. Then, the expansion is in powers of the momentum (mass) of the
pseudoscalar meson and the residue momentum of the heavy baryon. The chiral power counting is recovered. The HBChPT formalism has been used to study
the electromagnetic properties of the octet and decuplet baryons
\cite{Li:2016ezv,Napsuciale:1996ny,Meissner:1997hn,Jenkins:1992pi}. The masses of the charmed baryons are large, which are about $2.5$ GeV. In this case, the recoil effect is negligible. The heavy baryon chiral perturbation theory is suitable in calculating the magnetic moments of the heavy baryons.

 In Refs.~\cite{Tiburzi:2004mv,Banuls:1999mu}, the authors have calculated the magnetic moments of the heavy baryons up to next-to-leading order (NLO) using the (partially quenched) heavy hadron chiral perturbation theory. The wave-function renormalization, the vertex renormalization, and other effects are not included up to this order. However, they may not  be negligible. For instance, the wave-function renormalization contributes a nonanalytic $m_{q}\text{ln} m_{q}$ ($m_{q}$ is the light quark mass.) correction to the magnetic moments of the baryons~\cite{Jenkins:1992pi,Meissner:1997hn}. To include the above effects, we have considered all the one-loop diagrams. And we calculate the analytical expressions of the magnetic moments up to the next-to-next-to-leading order (NNLO).  Several relations of magnetic moments were given up to NLO in Ref.~\cite{Banuls:1999mu}. We find most of these relations are not valid any more at NNLO.

The paper is arranged as follows. In Sec.~\ref{sec1}, we give the
effective Lagrangians that contribute to the magnetic moments. In
Sec.~\ref{sec2}, we calculate the analytical expressions of the magnetic moments of the
antitriplet and the sextet charmed baryons. In Sec.~\ref{sec3}, we obtain the numerical results. The magnetic moments of antri-triplet charmed baryons are given to $\mathcal{O}(p^2)$. The numerical results of sextet charmed baryons are given up to $\mathcal{O}(p^3)$ in two scenarios.
The last section is a brief summary. Finally, some calculation details and explicit loop
integrals are collected in the Appendix.

\section{The Effective Lagrangians }\label{sec1}

\begin{figure}[tb]
 \centering
 \includegraphics[width=0.5\hsize]{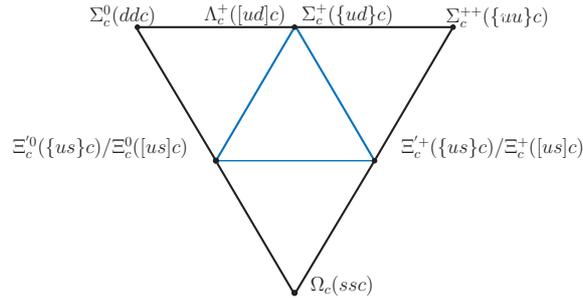}\\
 \caption{The blue and black triangles denote the heavy baryons in the $\bar 3_f$ and the $6_f$ flavor representations, respectively. $[...]$ and $\{... \}$ denotes the two light quarks are antisymmetric and symmetric in the flavor space, respectively.}\label{w2}
 \end{figure}
In the SU(3) flavor symmetry, the two light quarks in the heavy
baryon form the antisymmetric $\bar 3_f$ and the symmetric $6_f$
representations as illustrated in Fig.~\ref{w2}. The total spin of
the light quarks is $S_l=0$ or $S_l=1$, respectively. The spin of
the antitriplet heavy baryon is $S_{{\bar 3}}=\frac{1}{2}$ and the
spin of the sextet heavy baryon is $\frac{1}{2}$ or $\frac{3}{2}$.
We denote the above three kinds of states as $\psi_{{\bar 3}}$,
$\psi_{6}$ and $\psi^{*\mu}_{6}$ respectively.
\begin{eqnarray}
\label{wl1} \quad \psi_{{\bar{3}}}=\left(\begin{array}{ccc}
0 & \Lambda_{c}^{+} & \Xi_{c}^{+}\\
-\Lambda_{c}^{+} & 0 & \Xi_{c}^{0}\\
-\Xi_{c}^{+} & -\Xi_{c}^{0} & 0
\end{array}\right),\quad \psi_{{6}}=\left(\begin{array}{ccc}
\Sigma_{c}^{++} & \frac{\Sigma_{c}^{+}}{\sqrt{2}} & \frac{\Xi_{c}^{\prime+}}{\sqrt{2}}\\
\frac{\Sigma_{c}^{+}}{\sqrt{2}} & \Sigma_{c}^{0} & \frac{\Xi_{c}^{\prime0}}{\sqrt{2}}\\
\frac{\Xi_{c}^{\prime+}}{\sqrt{2}} &
\frac{\Xi_{c}^{\prime0}}{\sqrt{2}} & \Omega_{c}^{0}
\end{array}\right),\quad
\psi_{6}^{*\mu}=\left(\begin{array}{ccc}
\Sigma_{c}^{*++} & \frac{\Sigma_{c}^{*+}}{\sqrt{2}} & \frac{\Xi_{c}^{*+}}{\sqrt{2}}\\
\frac{\Sigma_{c}^{*+}}{\sqrt{2}} & \Sigma_{c}^{*0} & \frac{\Xi_{c}^{*0}}{\sqrt{2}}\\
\frac{\Xi_{c}^{*+}}{\sqrt{2}} & \frac{\Xi_{c}^{*0}}{\sqrt{2}} &
\Omega_{c}^{*0}
\end{array}\right)^{\mu}.
\end{eqnarray}
In the HBChPT scheme, we decompose the heavy baryon fields into the
``heavy" and ``light" components as follows

\begin{eqnarray}
\label{wl2b}
B_n(x)=e^{iM_B v\cdot x}\frac{1+\slashed v}{2}\psi_n ,~~H_n(x)=e^{iM_B v\cdot x}\frac{1-\slashed v}{2}\psi_n,
\end{eqnarray}
where $\psi_n$ denotes the heavy baryon field $\psi_{{\bar 3}}$,
$\psi_{6}$ or $\psi^{*\mu}_{6}$, and $B_n$ ($H_n$) is the ``light"
(``heavy'') component of the corresponding heavy baryon field. $M_B$
is the baryon mass, and $v^{\mu}=(1,\vec 0)$ is the static velocity.
The heavy field $H_n(x)$ is then integrated out in the Lagrangians.

In the HBChPT scheme, the matrix element of the electromagnetic
current for the spin-$\frac{1}{2}$ heavy baryon is

\begin{eqnarray}\label{wl10}
\langle B_{\bar3}(p')|\mathcal{J}_{\mu}|B_{\bar3}(p)\rangle=\bar u(p')\left(
v_{\mu}G_E(q^2)+\frac{[S_{\mu},S\cdot q]}{M_B}G_M(q^2)\right)u(p),
\end{eqnarray}
where $\mathcal{J}_{\mu}$ is the electromagnetic current, and
$q=p'-p$ is the transferred momentum. $u(p)$ and $\bar u(p)$
represent the Dirac spinors for the initial and finial heavy baryons
with the momentum $p$, respectively. $S^{\mu}$ is the spin operator
$\frac{i}{2}\gamma^5\sigma ^{\mu\nu}v_{\nu}$. $G_E$ and $G_M$ are
the electric and magnetic form factors, respectively. The magnetic
moment $\mu_B=\frac{e}{2M_B}G_M(0)$.

\subsection{The leading-order Lagrangians}
The pseudoscalar mesons are denoted as
\begin{eqnarray}
\phi=\sqrt{2}\left(\begin{array}{ccc}
\frac{\pi^{0}}{\sqrt{2}}+\frac{\eta}{\sqrt{6}} & \pi^{+} & K^{+}\\
\pi^{-} & -\frac{\pi^{0}}{\sqrt{2}}+\frac{\eta}{\sqrt{6}} & K^{0}\\
K^{-} & \overline{K}^{0} & -\frac{2}{\sqrt{6}}\eta
\end{array}\right) .
\end{eqnarray}
The Lagrangian for the pseudoscalar mesons at the leading order is
\begin{eqnarray}
\label{wl2}
&\mathcal{L}^{(2)}=\frac{F^2_0}{4}\text{Tr}[\triangledown_{\mu}U{\triangledown^{\mu}U}^{\dagger}],
\end{eqnarray}
with
\begin{eqnarray}
U=\xi^2=e^{\frac{i\phi}{F_0}}, \quad
&\triangledown_{\mu}U=\partial_{\mu} U+i eA_{\mu}[Q_l,U] ,
\end{eqnarray}
where $F_0$ is the decay
constant in the chiral limit. In this work, we use the values
$F_{\pi}=92.4$ MeV, $F_{K}=113$ MeV, and $F_{\eta}=116$ MeV after renormalization,
respectively. $A_\mu$ is the electromagnetic field. The charge matrix for the light quark is
$Q_l=\text{diag}(2/3,-1/3,-1/3)$.

The leading-order Lagrangians related to the heavy baryons are
\begin{eqnarray}\label{wl3}
\mathcal{L}_0^{(1)}&=&\frac{1}{2}\text{Tr}[{\bar \psi}_{{\bar{3}}}(i \slashed D-M_{{\bar{3}}})\psi_{{\bar{3}}}]+\text{Tr}[{\bar \psi}_{6}(i\slashed D-M_{6})\psi_{6}]\nonumber\\
&&+\text{Tr}\left[{\bar \psi}_{6}^{*\mu}(-g_{\mu\nu}(i\slashed D-M_{6^*})+i(\gamma_\mu D_\nu+\gamma_\nu D_\mu)-\gamma_\mu(i\slashed D+M_{6^*})\gamma_\nu)\psi_{6}^{*\nu}\right ]\\
\mathcal{L}_{\textrm{int}}^{(1)}&=&g_1\textrm{Tr}({\bar \psi}_{6}\slashed u\gamma_5\psi_{6})+g_2\textrm{Tr}[({\bar \psi}_{6}\slashed u\gamma_5\psi_{{\bar{3}}})+\textrm{H.c.}]+g_3\textrm{Tr}[({\bar \psi}_{6}^{*\mu}u_\mu \psi_{6})+\textrm{H.c.}]\nonumber\\
&&+g_4\textrm{Tr}[({\bar \psi}_{6}^{*\mu} u_\mu
\psi_{{\bar{3}}})+\textrm{H.c.}]+g_5\textrm{Tr}({\bar
\psi}_{6}^{*\nu}\slashed u\gamma_5 \psi_{6\nu}^{*}
)+g_6\textrm{Tr}({\bar \psi}_{{\bar{3}}}\slashed
u\gamma_5\psi_{{\bar{3}}}).
\end{eqnarray}
with the covariant derivatives and $u_{\mu}$ defined as
\begin{align*}
&D_{\mu}\psi_B=\partial_{\mu}\psi_B+\Gamma_{\mu}\psi_B+\psi_B\Gamma_{\mu}^{T}, \nonumber \\
&\Gamma_{\mu}=\frac{1}{2}(\xi^{\dagger}\partial_{\mu}\xi+\xi\partial_{\mu}\xi^{\dagger} )+\frac{i}{2}eA_{\mu}(\xi^{\dagger}Q_B \xi+\xi Q_B\xi^{\dagger} ),\nonumber \\
&u_{\mu}=\frac{i}{2}(\xi^{\dagger}\partial_{\mu}\xi-\xi\partial_{\mu}\xi^{\dagger}
)-\frac{1}{2}eA_{\mu}(\xi ^{\dagger}Q_B \xi-\xi Q_B\xi^{\dagger} ),
\end{align*}
where the charge operator of the heavy baryon $Q_{B}=\text{diag}(1,0,0)$. In this work, the mass difference
among the antitriplet multiplet and the sextet multiplet is
neglected. We use the average masses $M_{_{\bar{3}}}=2408$ MeV,
$M_{_6}=2535$ MeV, and $M_{6^*}=2602$ MeV, respectively
\cite{Patrignani:2016xqp}.

$g_i$ is the coupling for the interaction between the pseudoscalar
mesons and heavy baryons. $g_{2,4}$ are calculated through the
widths of the heavy baryons. $g_{1,3,5}$ are related to $g_{2,4}$
with the help of the quark model. Their values are
\cite{Yan:1992gz,Jiang:2014ena,Jiang:2015xqa}
\begin{eqnarray}
&g_1=0.98,\quad g_2=-\sqrt{\frac{3}{8}}g_1=-0.60,\quad
g_3=\frac{\sqrt{3}}{2}g_1=0.85, \nonumber\\
&g_4=-\sqrt{3}g_2=1.04,\quad g_5=-\frac{3}{2}g_1=-1.47, \quad g_6=0.
\end{eqnarray}
The pseudoscalar mesons only interact with the light quarks inside
the heavy baryons, and the total spin of light quarks is $S_l=0$ for
the baryon in the antitriplet. The $\phi B_{\bar 3} B_{\bar 3}$ vertex is
therefore forbidden considering the parity and angular momentum
conservation, and thus $g_6=0$.

The nonrelativistic Lagrangian at the leading order can be directly derived
from the $\mathcal{L}_0^{(1)}+\mathcal{L}_{int}^{(1)} $
\cite{Yan:1992gz, Jiang:2014ena},
\begin{eqnarray}\label{wl4}
\mathcal{L}_{MB}^{(1)}&=&\frac{1}{2}\text{Tr}(\bar{B}_{\bar{3}}iv\cdot DB_{\bar{3}})+\text{Tr}(\bar{B}_{6}(iv\cdot D-\delta_{1})B_{6})-\text{Tr}(\bar{B}_{6}^{*\mu}(iv\cdot D-\delta_{2})B_{6\mu}^{*})\nonumber\\
&& +2g_{1}\text{Tr}(\bar{B}_{6}S\cdot uB_{6})+2g_{2}\text{Tr}(\bar{B}_{6}S\cdot uB_{\bar{3}}+{\rm H.c.})+g_{3}\text{Tr}(\bar{B}_{6\mu}^{*}u^{\mu}B_{6}+{\rm H.c.})\nonumber\\
 && +g_{4}\text{Tr}(\bar{B}_{6\mu}^{*}u^{\mu}B_{\bar{3}}+{\rm H.c.})+2g_{5}\text{Tr}(\bar{B}_{6}^{*\mu}S\cdot uB_{6\mu}^{*})+2g_{6}\text{Tr}(\bar{B}_{\bar{3}}S\cdot uB_{\bar{3}}),
\end{eqnarray}
where the mass differences are $\delta_1=M_{6}-M_{\bar{3}}=127$ MeV,
$\delta_2=M_{{6^*}}-M_{\bar{3}}=194$ MeV and
$\delta_3=M_{{6^*}}-M_{6}=67$ MeV.

\subsection{The $\mathcal O(p^2)$ Lagrangians}

The nonrelativistic Lagrangians at $\mathcal{O}(p^2)$ contribute to the leading-order magnetic moments at the tree level:
\begin{eqnarray}\label{wl5}
&&{\mathcal {L}}^{(2)}_{33}=-\frac{id_2}{8M_N}\text{Tr}(\bar B_{\bar{3}}[S^{\mu},S^{\nu}]f_{\mu\nu}^{+}B_{\bar{3}})-\frac{id_3}{8M_N}\text{Tr}(\bar B_{\bar{3}}[S^{\mu},S^{\nu}]B_{\bar{3}})\text{Tr}(f_{\mu\nu}^{+}),\nonumber \\
&&{\mathcal
{L}}^{(2)}_{66}=-\frac{id_{5}}{4M_N}\text{Tr}(\bar{B_{6}}[S^{\mu},S^{\nu}]f_{\mu\nu}^{+}B_{6})
-\frac{id_{6}}{4M_N}\text{Tr}(\bar{B_{6}}[S^{\mu},S^{\nu}]B_{6})\text{Tr}(f_{\mu\nu}^{+}),\nonumber \\
&& {\mathcal
{L}}^{(2)}_{6^*6^*}=-\frac{id_{8}}{2M_N}\text{Tr}(\bar{B}_{6^*}^{\mu}f_{\mu\nu}^{+}{B}_{6^*}^{\nu})-\frac{id_{9}}{2
M_N}\text{Tr}(\bar{B}_{6^*}^{\mu}{B}_{6^*}^{\nu})\text{Tr}(f_{\mu\nu}^{+}),
\end{eqnarray}
where we use the subscript $6^{*}$ to indicate the spin-$3\over 2$
sextet. $M_N$ is the nucleon mass. The tensor
field $f^+_{\mu\nu}$ is defined as
\begin{eqnarray}\label{wl6}
&f^{\pm}_{\mu\nu}=\xi^{\dagger}f^R_{\mu\nu}\xi \pm \xi^{\dagger}f^L_{\mu\nu}\xi,\nonumber\\
&f^R_{\mu\nu}=f^L_{\mu\nu}=-eQ_B(\partial_{\mu}A_{\nu}-\partial_{\nu}A_{\mu}).
\end{eqnarray}
$f^+_{\mu\nu}$ belongs to the $1\oplus8$ representation in the
flavor space. For the heavy baryons, $\bar 3\otimes 3=1\oplus8$,
$\bar 6 \otimes 6 =1 \oplus8 \oplus27$, and thus the Lagrangians are
constructed in two ways, $8\otimes 8 \rightarrow 1$ and $1\otimes
1\rightarrow 1$. Therefore, each Lagrangain in Eq. (\ref{wl5})
contains two independent interaction terms. The contributions to the
magnetic moments from the $d_{2,5,8}$ terms are proportional to the
total charges of the heavy baryons. The $d_{3,6,9}$ terms represent
the contributions from the heavy quark since $\text{Tr}(Q_l)=0$. The
$d_8$ and $d_9$ terms contribute to the magnetic moments of the
spin-$1 \over 2$ heavy baryons at $\mathcal{O}(p^3)$ through the
loop diagrams.

To calculate the magnetic moments up to $\mathcal O(p^3)$, we also
need the leading-order Lagrangians which directly contribute to the
transition magnetic moments,
\begin{eqnarray}\label{wl7}
&&{\mathcal {L}}^{(2)}_{6\rightarrow 3}=-\frac{2if_2}{M_N}\text{Tr}(\bar B_{\bar 3}f^+_{\mu\nu}[S^{\mu},S^{\nu}]B_6)+\textrm{H.c.},\nonumber\\
&&{\mathcal {L}}^{(2)}_{6^*\rightarrow 3}=-i\frac{f_4}{2M_N}\text{Tr}(\bar B_6^{*\mu} f^{+}_{\mu\nu}S^{\nu}B_{\bar 3})+\textrm{H.c.},\nonumber\\
&&{\mathcal {L}}^{(2)}_{6^*\rightarrow
6}=-i\frac{f_3}{2M_N}\text{Tr}(\bar B_6^{*\mu}
f^{+}_{\mu\nu}S^{\nu}B_{6})-i\frac{\tilde{f_3}}{2M_N}\text{Tr}(\bar
B_6^{*\mu}S^{\nu}B_6)\text{Tr}(f^+_{\mu\nu})+\textrm{H.c.}
\end{eqnarray}
Since $3\otimes 6=8\oplus10$, only one independent Lagrangian term
contributes to the radiative transition between the antitriplet and
sextet. The ${\mathcal {L}}^{(2)}_{6^*\rightarrow 6}$ is similar to
the $\mathcal L_{66,6^{*}6^{*}}^{(2)}$ and has two independent
interaction terms. We denote them as the $f_3$ and $\tilde f_3$
terms in this work while the authors of Ref.~\cite{Jiang:2015xqa}
denote them as $\text{Tr}(\bar B_6\epsilon^{\mu \nu \rho
\sigma}f^+_{\mu\nu})iD_{\rho}B^*_{\sigma}$ and $\text{Tr}(\bar
B_6\epsilon^{\mu \nu \rho
\sigma}iD_{\rho}B^{\mu*}_{\sigma})\text{Tr}( f^+_{\mu\nu})$. These
two terms can be transformed into ${\mathcal
{L}}^{(2)}_{6^*\rightarrow 6}$ with the conditions $v \cdot B^*=0$,
$S\cdot B^*=0$ and $v\cdot S=0$.

The following $\mathcal O(p^2)$ Lagrangian for the pseudoscalar
mesons and the heavy baryons also contribute to the $\mathcal
O(p^3)$ magnetic moments through the vertex correction,
\begin{eqnarray}\label{wl8}
&&{\mathcal {L}}^{(2)}_{MB}=\frac{d_{1}}{2M_N}\text{Tr}(\bar
B_{\bar{3}}[S^{\mu},S^{\nu}][u_{\mu},u_{\nu}]B_{\bar{3}})+\frac{d_{4}}{M_N}{\Tr}(\bar{B}_{6}[S^{\mu},S^{\nu}][u_{\mu},u_{\nu}]B_{6}).
\end{eqnarray}
The other Lagrangian terms with the structures $\text{Tr}(\bar
B[S^{\mu},S^{\nu}]\{u_{\mu},u_{\nu}\}B)$ vanish due to the antisymmetric Lorentz indices of $[S^{\mu},S^{\nu}]$.

\subsection{The $\mathcal O(p^4)$ Lagrangians}

The ${\mathcal O}(p^3)$ Lagrangian does not contribute to the
magnetic moment up to the next-to-next-to-leading order. The
${\mathcal O}(p^4)$ heavy baryon Lagrangians contributing to the
magnetic moments read
\begin{eqnarray}\label{wl9}
&&{\mathcal L}_{66}^{(4)}=-\frac{is_{1}}{4M_N}\text{Tr}(\bar{B}_{6}[S^{\mu},S^{\nu}]\{ \chi_{+},f_{\mu\nu}^{+}\}B_{6})-\frac{is_{2}}{4M_N}\text{Tr}(\bar{B}_{6}[S^{\mu},S^{\nu}]\chi_{+}B_{6})\text{Tr}(f_{\mu\nu}^{+})\nonumber\\
&&~~~~~~~~~-\frac{is_{3}}{4M_N}\text{Tr}(\bar{B}_{6}[S^{\mu},S^{\nu}]f_{\mu\nu}^{+}B_{6}\chi_{+}^{T})-\frac{is_{4}}{4M_N}\text{Tr}(\bar{B}_{6}[S^{\mu},S^{\nu}]B_{6})\text{Tr}(\chi_{+}f_{\mu\nu}^{+}),\label{l466}\\
&&{\mathcal L}_{33}^{(4)}=-\frac{1}{2}\frac{is_{5}}{4M_N}\text{Tr}(\bar B_{\bar{3}}[S^{\mu},S^{\nu}]\{ \chi_{+},f_{\mu\nu}^{+}\}B_{\bar{3}})-\frac{1}{2}\frac{is_{6}}{4M_N}\text{Tr}(\bar B_{\bar{3}}[S^{\mu},S^{\nu}]\chi_{+}B_{\bar{3}})\text{Tr}(f_{\mu\nu}^{+})\nonumber\\
&&~~~~~~~~~-\frac{1}{2}\frac{is_{7}}{4M_N}\text{Tr}(\bar
B_{\bar{3}}[S^{\mu},S^{\nu}]B_{\bar{3}})\text{Tr}(\chi_{+}f_{\mu\nu}^{+}).\label{l433}
\end{eqnarray}
At this order, the effect of the SU(3) symmetry breaking is
introduced through the current quark mass matrix.
\begin{eqnarray}
&\chi_{\pm}=\xi^{\dagger}\chi\xi^{\dagger} \pm \xi \chi^{\dagger} \xi,\nonumber\\
&\chi=2B_0\text{Diag}(m^{c}_u,m^{c}_d,m^{c}_s),
\end{eqnarray}
where $B_0$ is a parameter related to the quark condensate.
$m^c_{u,d,s}$ is the mass of the light current quark. At the leading
order, $\chi_{+}=\text{diag}(0,0,1)$ if we assume $m^c_{u,d}=0$ and
absorb the $2B_0 m^c_s$ into the LECs $s_1\sim s_7$.

In general, there should exist six terms for
$\mathcal{L}_{33}^{(4)}$, which are listed in the first six columns
of Table~\ref{flovorstructure}. The terms with
$\text{Tr}(\chi_+)f_{\mu\nu}^+$ and
$\text{Tr}(\chi_{+})\text{Tr}(f_{\mu\nu}^+)$ can be absorbed in Eq.
(\ref{wl5}) by the redefinition of the LECs $d_2$ and $d_3$,
respectively. The leading term of the $8_1$ structure
$[\chi_{+},f_{\mu\nu}^{+}]$ is always $0$ after expansion. Thus,
there are only three independent terms in Eq. (\ref{l433}).
For the $\mathcal{L}_{66}^{(4)}$, an extra flavor structure in the
last column of Table~\ref{flovorstructure} is introduced because of
$6\otimes\bar{6}\rightarrow 27$, which corresponds to the $s_3$ term
in Eq. (\ref{l466}). Thus, there are four independent terms in
$\mathcal{L}_{66}^{(4)}$ in total.
\begin{table}
 \caption{The possible flavor structures which contribute to the $\mathcal{O}(p^3)$ magnetic moments at tree level. $(\chi_{+}f_{\mu\nu}^{+})_{ab}^{ij}\equiv(\chi_+)^{\{i}_{\{a}(f^+_{\mu\nu})^{j\}}_{b\}}$, where the $\{...\}$ means that the flavor scripts are symmetrized. }\label{flovorstructure}
\begin{tabular}{l|c|c|c|c|c|c|c}
\toprule[1pt]\toprule[1pt] Group representation &
$1\otimes1\rightarrow1$ & $1\otimes8\rightarrow8$ &
$8\otimes1\rightarrow8$ & $8\times8\rightarrow1$ &
$8\otimes8\rightarrow8_{1}$ & $8\otimes8\rightarrow8_{2}$ &
$8\otimes8\rightarrow27$\tabularnewline \midrule[1pt] Flavor
structure & $\text{Tr}(\chi_{+})\text{Tr}(f_{\mu\nu}^{+})$ &
$\text{Tr}(\chi_{+})f_{\mu\nu}^{+}$ &
$\chi_{+}\text{Tr}(f_{\mu\nu}^{+})$ &
$\text{Tr}(\chi_{+}f_{\mu\nu}^{+})$ & $[\chi_{+},f_{\mu\nu}^{+}]$ &
$\{\chi_{+},f_{\mu\nu}^{+}\}$ &
$(\chi_{+}f_{\mu\nu}^{+})_{ab}^{ij}$\tabularnewline
\bottomrule[1pt]\bottomrule[1pt]
\end{tabular}
\end{table}

\section{The magnetic moments of the spin-$1\over 2$ heavy baryons}\label{sec2}
\begin{figure}[tb]
 \centering
 \includegraphics[scale=1.0]{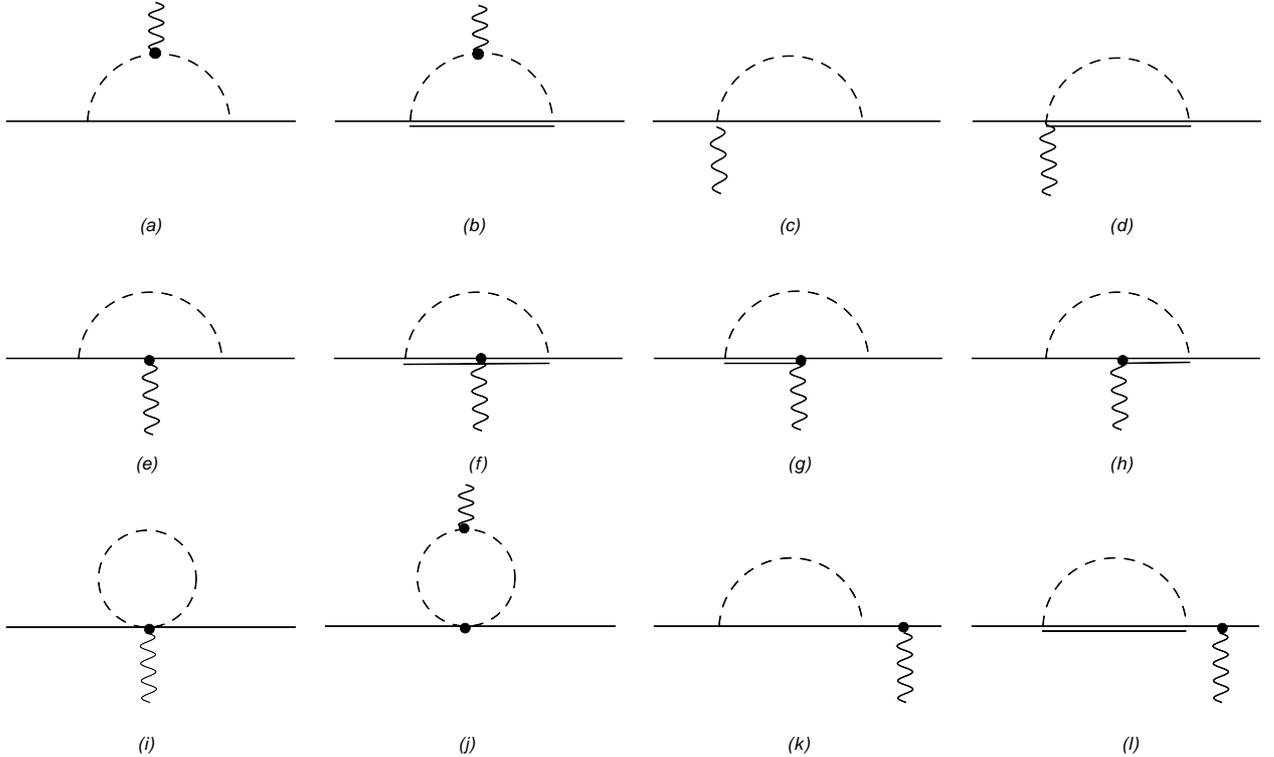}
 \caption{The loop diagrams contribute to the magnetic moments of the heavy baryons up to $\mathcal{O}(p^3)$. The solid dot represents the $\mathcal{O}(p^2)$ vertex. The single and double lines denote the spin-$1\over 2$ and spin-$3\over 2$ heavy baryons, respectively.}\label{all loop}
 \end{figure}

We list the loop diagrams contributing to $\mu_B$ up to the
next-to-next-to-leading order in Fig.~\ref{all loop}. A diagram with
chiral dimension $D_{\chi}$ contributes to the magnetic moments at
$D_{\chi}-1$ order.

\subsection{The magnetic moments of the heavy baryons in the $\bar 3_f$ flavor representation.}\label{subsec1}
The magnetic moments at the leading order are derived from the Lagrangians
in Eq. (\ref{wl5}),
\begin{eqnarray}\label{op13bar}
\mu^{(1)}_{\Lambda_{c}^{+}}=\frac{1}{2}\left(d_{2}+2d_{3}\right), &
\mu^{(1)}_{\Xi_{c}^0}=d_{3},
&\mu^{(1)}_{\Xi_{c}^{+}}=\frac{1}{2}\left(d_{2}+2d_{3}\right).
\end{eqnarray}

For the $\bar 3_f$ heavy baryons, the
intermediate baryons in the loops contain only the baryons in the
$6_f$ representation since $g_6=0$.

At $\mathcal{O}(p^2)$, the chiral correction to the magnetic moments
comes from diagrams (a) and (b) in Fig.~\ref{all loop} and is written as,
\begin{eqnarray}
\label{32}
&&\mu^{(2;a,b)}=-(\frac{g_{2}}{F_x})^{2}\beta^{x}M_{N}n_{1}^{\textrm{II}}(-\delta_{1},m_{x})+(\frac{g_{4}}{2F_x})^{2}\beta^{x}M_{N}\frac{4}{d-1}n_{1}^{\textrm{II}}(-\delta_{2},m_{x}),
\end{eqnarray}
where the superscript $(2;a,b)$ denotes the chiral order and the Feynman diagrams. The script $x$ indicates the pseudoscalar mesons in the loops. $m_x$ and $F_x$ are their masses and decay constants in the chiral limit. $d$ is the dimension. $\beta^x$ is the Clebsch-Gordan coefficient for the heavy baryons as listed in Table~\ref{magnetic3}. The notations such as the $n_{1}^{\textrm{II}}$ and $J_2(w)$ in Eq.(\ref{wl11}) are the loop integrals. Their expressions are collected in Appendix
\ref{app1}. The
chiral corrections from diagrams $(c)$ and $(d)$ vanish since
their amplitudes contain the structure $S\cdot v=0$ after the loop
integration.

At $\mathcal{O}(p^3)$, the loop contributions are from
diagrams (e)-(l).
\begin{eqnarray}
\label{wl11}
\mu^{(3;e-h)}&=&\frac{g_{2}^2}{F_x^2}\theta_{1}^{x}\frac{3-d}{4}J'_2(-\delta_{1})
-2\frac{g_{2}g_{4}}{F_x^{2}}\theta_{3}^{x}\frac{3-d}{d-1}\frac{J_{2}(-\delta_{1})-J_{2}(-\delta_{2})}{\delta_{1}-\delta_{2}}\nonumber \\
&&-\frac{g_{4}^2}{4F_x^2}\theta_{2}^{x}\left(\frac{8}{1-d}+\frac{4(5-d)}{(d-1)^{2}}\right)J'_2(-\delta_{2}),
\nonumber \\
\mu^{(3;i,j)}&=&\frac{\delta^x}{4F_x^{2}}m_{x}^{2}\frac{1}{16\pi^{2}}\textrm{ln}\frac{m_{x}}{\lambda}+\frac{\alpha^{x}}{F_x^{2}}m_{x}^{2}\frac{1}{16\pi^{2}}\textrm{ln}\frac{m_{x}}{\lambda},\nonumber\\
\mu^{(3;k,l)}&=&\frac{g_{2}^2}{F_x^2}N^{x}\frac{1-d}{4}J'_2(-\delta_{1})\mu_{3}^{(1)}+
\frac{g_{4}^2}{4F_x^2}N^{x}(2-d)J'_2(-\delta_{2})\mu_{3}^{(1)},
\end{eqnarray}
where $\lambda$ is the cut off parameter and we adopt $\lambda=$1GeV in this work. $\mu_{3}^{(1)}$ is the
magnetic moment at the leading order in Eq. (\ref{op13bar}). In
Table~\ref{magnetic3}, we list the coefficients $\theta^x$,
$\delta^x$, and so on.
The $\mathcal O(p^3)$ magnetic moment $\mu^{(3;tree)}$ from the tree
diagram reads,
 \begin{eqnarray}
\mu^{(3;tree)}_{\Lambda_{c}^{+}}=0,
~\mu^{(3;tree)}_{\Xi^0_{c}}=\frac{1}{2} s_{6},
~\mu^{(3;tree)}_{\Xi_{c}^{+}}=\frac{1}{2}s_6.
\end{eqnarray}

\begin{table}
 \caption{The coefficients in Eq. (\ref{wl11}) for the magnetic moments of the heavy baryons in $\bar 3_f$.}\label{magnetic3}
\begin{tabular}{c|c|c|c|c}
\toprule[1pt]\toprule[1pt]
 Loop& \multicolumn{1}{c|}{} & \multicolumn{1}{c|}{$\Lambda_{c}^{+}$} & \multicolumn{1}{c|}{$\Xi_{c}^0$} & $\Xi_{c}^{+}$\tabularnewline
\midrule[1pt] \multirow{2}{*}{(a),(b)} &
\multicolumn{1}{c|}{$\beta^{\pi}$} & $0$ & $-1$ &
$\ensuremath{1}$\tabularnewline
 & \multicolumn{1}{c|}{$\beta^{k}$} & $1$ & $-1$ & $\ensuremath{0}$\tabularnewline
\hline \hline \multirow{4}{*}{(e)-(h)} &
\multicolumn{1}{c|}{$\theta_{1}^{\pi}$} &
$3\left(d_{5}+2d_{6}\right)$ &
$\frac{1}{2}\left(d_{5}+3d_{6}\right)$ &
$\frac{1}{4}\left(d_{5}+6d_{6}\right)$\tabularnewline
 & \multicolumn{1}{c|}{$\theta_{1}^{k}$} & $\frac{1}{2}\left(d_{5}+4d_{6}\right)$ & $\frac{d_{5}}{2}+5d_{6}$ & $\frac{5}{2}d_{5}+5d_{6}$\tabularnewline
 & \multicolumn{1}{c|}{$\theta_{1}^{\eta}$} & $0$ & $\frac{3d_{6}}{2}$ & $\frac{3}{4}\left(d_{5}+2d_{6}\right)$\tabularnewline
\cline{2-5}
 & \multicolumn{4}{c}{$\theta_{2}=\theta_{1}(d_{5}\rightarrow d_{8},d_{6}\rightarrow d_{9})$
, $\theta_{3}=\theta_{1}(d_{5}\rightarrow f_{3},d_{6}\rightarrow\tilde{f_{3}})$}\\
\hline \multirow{3}{*}{(k),(l)} & $N^{\eta}$ & $0$ & $\frac{3}{2}$ &
$\frac{3}{2}$\tabularnewline
 & $N^{\pi}$ & $6$ & $\frac{3}{2}$ & $\frac{3}{2}$\tabularnewline
 & $N^{k}$ & $2$ & $5$ & $5$\tabularnewline
\hline \multirow{2}{*}{(i)} & $\delta^{\pi}$ & $0$ & $2d_{2}$ &
$-2d_{2}$\tabularnewline
 & $\delta^{k}$ & $-2d_{2}$ & $2d_{2}$ & $0$\tabularnewline
\hline \multirow{2}{*}{(j)} & $\alpha^{\pi}$ & $0$ & $-d_{1}$ &
$d_{1}$\tabularnewline
 & $\alpha^{k}$ & $d_{1}$ & $-d_{1}$ & $0$\tabularnewline
\bottomrule[1pt]\bottomrule[1pt]
\end{tabular}
\end{table}

\subsection{The magnetic moments of the heavy baryons in the $6_f$ flavor representation.}\label{subsec2}

For the spin-$1\over 2$ heavy baryons in the $6_f$ representation,
their magnetic moments at the leading order are
\begin{eqnarray}
\mu_{\Sigma_{c}^{++}}^{(1)}=d_{5}+d_{6},~
\mu_{\Sigma_{c}^{+}}^{(1)}= \frac{d_{5}}{2}+d_{6},~
\mu_{\Sigma_{c}^{0}}^{(1)}={d_{6}},~\mu_{^{\Xi_{c}'^{+}}}^{(1)}=\frac{d_{5}}{2}+d_{6},~\mu_{^{\Xi_{c}'}}^{(1)}={d_{6}},~\mu_{\varOmega_{c}}^{(1)}=\ensuremath{{d_{6}}},\label{op16}
\end{eqnarray}
The $\mathcal O(p^2)$ and $\mathcal O(p^3)$ loop corrections are
listed as follows,
\begin{eqnarray}\label{wl12}
\mu^{(2;a,b)}&=&-\frac{g_{1}^2}{F_x^2}\beta^{x}M_N n_{1}^{\textrm{II}}(0,m_{x})+\frac{g_{3}^2}{4F_x^2}\frac{4}{d-1}\beta^{x}M_Nn_{1}^{\textrm{II}}(-\delta_{3},m_{x})-\frac{g_{2}^2}{F_x^{2}}h^{x}M_Nn_{1}^{\textrm{II}}(\delta_{1},m_{x}),\nonumber\\
\mu^{(3;k,l)}&=&\frac{g_{1}^2}{F_x^2}O^{x}\frac{1-d}{4}J'_2(0)\mu_{6}^{(1)}+\frac{g_{2}^2}{F_x^2}N^{x}\frac{1-d}{4}J'_2(\delta_{1})\mu_{6}^{(1)}
+\frac{g_{3}^2}{4F_x^2}O^{x}(2-d)J'_2(-\delta_{3})\mu_{6}^{(1)},\nonumber\\
\mu^{(3;e-h)}&=&\frac{g_{1}^2}{F_x^2}\rho_1^{x}\frac{3-d}{4}J'_2(0)+\frac{g_{2}^2}{F_x^2}\theta^{x}\frac{3-d}{4}J'_2(\delta_{1})+X^{x}\frac{8f_{2}g_{1}g_{2}}{F_x^{2}}\frac{d-3}{2}\Gamma_{2}(\delta_{1})-2\frac{g_{1}g_{3}}{F_x^{2}}\rho_{3}^{x}\frac{3-d}{d-1}\Gamma_2(-\delta_{3})\nonumber\\
&&~~-\frac{g_{3}^{2}}{4F_x^2}\rho_{2}^{x}\left(\frac{8}{1-d}+\frac{4(5-d)}{(d-1)^{2}}\right)J'_2(-\delta_{3})-2\frac{g_{2}g_{3}f_4}{F_x^{2}}X^{x}\frac{3-d}{d-1}\frac{J_{2}(\delta_{1})-J_{2}(-\delta_{3})}{-\delta_{3}-\delta_{1}},\nonumber\\
\mu^{(3;i,j)}&=&\frac{\alpha^{x}}{F_x^{2}}m_{x}^{2}\frac{1}{32\pi^{2}}\textrm{ln}\frac{m_{x}^{2}}{\lambda^{2}}+\frac{\delta^{x}}{8F_x^{2}}m_{x}^{2}\frac{1}{8\pi^{2}}\textrm{ln}\frac{m_{x}}{\lambda},
\end{eqnarray}
where $\mu_{6}^{(1)}$ is the leading-order magnetic moment in Eq.
(\ref{op16}). $\Gamma_2$ is another loop integral which is given
in Appendix~\ref{app1}. The relevant coefficients for the heavy
baryons in the sextet, such as $\beta^x$ and $O^x$, are listed in
Table~\ref{6}.
At $\mathcal{O}(p^3)$, the magnetic moments from the tree diagram are
\begin{eqnarray}
~ \mu_{\Sigma_c^{++}}^{(3;tree)}=0, ~\mu_{\Sigma_c^{+}}^{(3;tree)}=0
, ~\mu_{\Sigma_c^{0}}^{(3;tree)}=0,
~\mu_{^{\Xi'^{+}_c}}^{(3;tree)}=\frac{s_2}{2}+\frac{s_3}{2},
~\mu_{^{\Xi'^{0}_c}}^{(3;tree)}=\frac{s_2}{2},~\mu_{\Omega_c}^{(3;tree)}=s_2.\nonumber
\end{eqnarray}

\subsection{Analytical relations}
If we take the same mass for light quarks, strong interaction and electromagnetic interaction can not distinguish $d$ from $s$ quark, which gives rise to the U-spin symmetry. In the U-spin transformation, the quarks transform as $d\leftrightarrow s$, $\bar{d}\leftrightarrow \bar{s}$. The pseduscalar mesons tranfrom as $\pi^{\pm} \leftrightarrow K^{\pm}$ and $K^0\leftrightarrow \bar{K}^0$. The singly charmed baryons transform as $\Lambda_c^+\leftrightarrow \Xi_c^+$, $\Xi_c^0\leftrightarrow \Xi_c^0$, ${\Omega^0_c} \leftrightarrow{\Sigma^0_c}$, ${\Sigma^+_c}\leftrightarrow {\Xi^{'+}_c}$ and $\Xi_c^{'0}\leftrightarrow \Xi_c^{'0}$. Then the U-spin symmetry
leads to relations between the corresponding coefficients of
different heavy baryons. In the leading-order results, 
\begin{eqnarray}
\mu^{(1)}_{\Lambda_{c}^{+}}=\mu^{(1)}_{\Xi_{c}^{+}},\quad \mu_{\Sigma_{c}^{+}}^{(1)}= \mu_{^{\Xi_{c}'^{+}}}^{(1)},\quad \mu_{\Sigma_{c}^{0}}^{(1)}=\mu_{\varOmega_{c}}^{(1)}
\end{eqnarray}

In the diagrams $(a)$, $(b)$, $(i)$ and $(j)$, the photon only interacts with the charged pseudoscalar mesons, $\pi^{\pm}$ and $K^{\pm}$. The coefficients are related to each other as follows,
\begin{eqnarray}
\beta
^{\pi(K)}_{\Lambda^+_c}=\beta ^{K(\pi)}_{\Xi^+_c},~~\beta
^{\pi(K)}_{\Xi^0_c}=\beta ^{K(\pi)}_{\Xi^0_c},~~
\beta
^{\pi(K)}_{\Omega^0_c}=\beta ^{K(\pi)}_{\Sigma^0_c},~~\beta
^{\pi(K)}_{\Sigma^+_c}=\beta ^{K(\pi)}_{\Xi^{'+}_c},~~\beta
^{\pi(K)}_{\Xi^{'0}_c}=\beta ^{K(\pi)}_{\Xi^{'0}_c}.
\end{eqnarray}
 The $h$,$\theta $, $\alpha$ and the $\delta$ also obey the similar relations. In fact, this conclusion also applies in the (e)-(f), (k) and (l) diagrams with the charged intermediate pseudoscalar mesons in the loops, i.e. 
\begin{eqnarray}
N
^{\pi^\pm(K^\pm)}_{\Lambda^+_c}=N ^{K^\pm(\pi^\pm)}_{\Xi^+_c},~~N
^{\pi^\pm}_{\Xi^0_c}=N ^{K^\pm}_{\Xi^0_c},~~
N
^{\pi^\pm(K^\pm)}_{\Omega^0_c}=N ^{K^\pm(\pi^\pm)}_{\Sigma^0_c},~~N
^{\pi^\pm(K^\pm)}_{\Sigma^+_c}=N ^{K^\pm(\pi^\pm)}_{\Xi^{'+}_c},~~N
^{\pi^\pm}_{\Xi^{'0}_c}=N ^{K^\pm}_{\Xi^{'0}_c}.
\end{eqnarray}
where the coefficients $N$ can be replaced by the coefficients $O$, $\rho$, $\theta$ or $X$ in Tables \ref{magnetic3} and ~\ref{6}. 
 
When we neglect the mass splitting of pseudoscalar mesons and ignore the explicit SU(3) breaking terms in the Lagrangian, the interaction can not distinguish the $d$ from $s$ quark. Then the baryons with the same charge have the same magnetic moments at every order,
\begin{equation}
\sum_{\phi}C^{\phi}_{\Xi_c^{+}}=\sum_{\phi}C^{\phi}_{\Lambda_c^{+}},\quad\sum_{\phi} C^{\phi}_{\Sigma_c^0}=\sum_{\phi}C^{\phi}_{\Xi_c^{'0}}=\sum_{\phi}C^{\phi}_{\Omega_c^{0}},\quad 
\sum_{\phi}C^{\phi}_{\Xi_c^{'+}}=\sum_{\phi}C^{\phi}_{\Sigma_c^{+}}
\end{equation}
where $C$ denotes the coefficients $X$, $h$, $O$, $\theta$, $\alpha$, $\delta$, $N$, $\rho$, $\theta$ or $\beta$ in Table~\ref{magnetic3} and~\ref{6}. 
 

At $\mathcal O(p^2)$, we can also derive some relations from Table~\ref{6}. 
\begin{eqnarray}\label{wr}
&\mu^{(2;a,b)}_{\Sigma_{c}^{++}}=-2\mu^{(2;a,b)}_{\Xi_{c}^{'0}}, ~~\mu^{(2;a,b)}_{\Omega_{c}^{0}}=-2\mu^{(2;a,b)}_{\Sigma_{c}^{+}},~~ \mu^{(2;a,b)}_{\Sigma_{c}^{0}}=-2\mu^{(2;a,b)}_{\Xi_{c}^{'+}},\nonumber\\
& \mu^{(2;a,b)}_{\Sigma_{c}^{++}}+\mu^{(2;a,b)}_{\Sigma_{c}^{0}}+\mu^{(2;a,b)}_{\Omega_{c}^{0}}=0, ~~\mu^{(2;a,b)}_{\Xi_{c}^{'+}}+\mu^{(2;a,b)}_{\Xi_{c}^{'0}}+\mu^{(2;a,b)}_{\Sigma_{c}^{+}}=0,
\end{eqnarray}
Considering the results in leading order, there are several relations up to $\mathcal O(p^2)$, which are same as those in Ref.~\cite{Banuls:1999mu},
\begin{eqnarray}
&\mu_{\Sigma_c^{++}}+\mu_{\Sigma^0_c}=2\mu_{\Sigma^+_c},\\
&\mu_{\Sigma_c^{++}}+\mu_{\Omega^0_c}=2\mu_{\Xi^{'+}_c},\\
&\mu_{\Sigma_c^{++}}+2\mu_{\Xi^{'0}_c}=\mu_{\Sigma^0_c}+2\mu_{\Xi^{'+}_c}=\mu_{\Omega_{c}^{0}}+2\mu_{\Sigma_{c}^{+}}.
\end{eqnarray}
Up to $\mathcal O(p^3)$, the first relation is still valid, while the other two relations do not hold any more. The self-energy diagrams (k)-(l) and the loops (e)-(h) at $\mathcal O(p^4)$ contribute to the $\mathcal O(p^3)$ magnetic moments and destroy the above two relations. 

\begin{table}
\caption{The coefficients in Eq. (\ref{wl12}) for the magnetic moments of the heavy
baryons in $6_f$.}\label{6}
\begin{tabular}{c|c|c|c|c|c|c|c}
\toprule[1pt]\toprule[1pt]
 Loop& & ${\Sigma_{c}^{++}}$ & ${\Sigma_{c}^{+}}$ & ${\Sigma_{c}^{0}}$ & ${{\Xi_{c}'^{+}}}$ & ${{\Xi_{c}'^0}}$ & ${\varOmega_{c}^0}$\tabularnewline
\midrule[1pt]
\multirow{4}{*}{(a),(b)} & $\beta^{\pi}$ & $1$ & $\ensuremath{0}$ & $\ensuremath{-1}$ & $\frac{{1}}{2}$ & $-\frac{1}{2}$ & $0$\\
 & $\beta^{k}$ & $1$ & $\ensuremath{\frac{1}{2}}$ & $\ensuremath{0}$ & $0$ & $-\frac{1}{2}$ & $-\ensuremath{1}$\tabularnewline
\cline{2-8}
 & $h^{\pi}$ & $2$ & $\ensuremath{0}$ & $\ensuremath{-2}$ & $1$ & $-1$ & $0$\\
 & $h^{k}$ & $2$ & $\ensuremath{1}$ & $0$ & $0$ & $-1$ & $-2$\tabularnewline
\hline \hline
\multirow{2}{*}{(i)} & $\ensuremath{\delta^{\pi}}$ & $-\ensuremath{4d_{5}}$ & $\ensuremath{0}$ & $\ensuremath{4d_{5}}$ & $\ensuremath{-2d_{5}}$ & $2d_{5}$ & $0$\\
 & $\delta^{k}$ & $-\ensuremath{4d_{5}}$ & $\ensuremath{-2d_{5}}$ & $0$ & $\ensuremath{0}$ & $2d_{5}$ & $\ensuremath{}4d_{5}$\tabularnewline
\hline
\multirow{2}{*}{(j)} & $\alpha^{\pi}$ & $2\ensuremath{d_{4}}$ & $\ensuremath{0}$ & $-2d_{4}$ & $d_{4}$ & $-d_{4}$ & $0$\\
 & $\alpha^{k}$ & $2\ensuremath{d_{4}}$ & $d_{4}$ & $0$ & $0$ & $-d_{4}$ & $-\ensuremath{2d_{4}}$\tabularnewline
\hline
\multirow{6}{*}{(k),(l)} & $\ensuremath{O^{\pi}}$ & $2$ & $\ensuremath{2}$ & $\ensuremath{2}$ & $\ensuremath{\frac{{3}}{4}}$ & $\frac{3}{4}$ & $0$\\
 & $\ensuremath{O^{k}}$ & $1$ & $\ensuremath{1}$ & $1$ & $\ensuremath{\frac{5}{2}}$ & $\frac{5}{2}$ & $2$\\
 & $\ensuremath{O^{\eta}}$ & $\ensuremath{\frac{1}{3}}$ & $\frac{1}{3}$ & $\ensuremath{\frac{1}{3}}$ & $\ensuremath{\frac{1}{12}}$ & $\frac{1}{12}$ & $\ensuremath{\frac{4}{3}}$\tabularnewline
\cline{2-8}
 & $N^{\pi}$ & $2$ & $\ensuremath{2}$ & $\ensuremath{2}$ & $\ensuremath{\frac{{3}}{2}}$ & $\frac{3}{2}$ & $0$\\
 & $\ensuremath{N^{k}}$ & $2$ & $\ensuremath{2}$ & $\ensuremath{2}$ & $\ensuremath{1}$ & $1$ & $4$\\
 & $\ensuremath{N^{\eta}}$ & $0$ & $\ensuremath{0}$ & $\ensuremath{0}$ & $\ensuremath{\frac{3}{2}}$ & $\frac{3}{2}$ & $\ensuremath{0}$\tabularnewline
\hline

\multirow{10}{*}{(e)-(h)} & $\rho_1^{\pi}$ & $\ensuremath{\frac{3d_{5}}{2}+2d_{6}}$ & $\ensuremath{d_{5}+2d_{6}}$ & $\ensuremath{\frac{1}{2}\left(d_{5}+4d_{6}\right)}$ & $\ensuremath{\frac{1}{8}d_{5}+\frac{3}{4}d_{6}}$ & $\frac{3}{4}d_{6}+\frac{1}{4}d_{5}$ & $0$\\
 & $\rho_1^{k}$ & $\ensuremath{\frac{d_{5}}{2}+d_{6}}$ & $\ensuremath{d_{6}+\frac{1}{4}d_{5}}$ & $\ensuremath{d_{6}}$ & $\ensuremath{\frac{{5}}{4}(d_{5}+2d_{6})}$ & $\frac{1}{4}\left(d_{5}+10d_{6}\right)$ & $\ensuremath{2d_{6}+\frac{d_{5}}{2}}$\\
 & $\rho_1^{\eta}$ & $\ensuremath{\frac{1}{3}\left(d_{5}+d_{6}\right)}$ & $\ensuremath{\frac{1}{3}(\frac{1}{2}d_{5}+d_{6})}$ & $\ensuremath{\frac{d_{6}}{3}}$ & $\ensuremath{\frac{1}{12}(\frac{1}{2}d_{5}+d_{6})}$ & $\frac{1}{12}d_{6}$ & $\ensuremath{\frac{4}{3}d_{6}}$\tabularnewline
\cline{2-8}
 & \multicolumn{7}{c}{$\ensuremath{\rho_{2}^{x}=\rho_{1}^{x}(d_{5}\rightarrow d_{8},d_{6}\rightarrow d_{9})},\ensuremath{\rho_{3}=\rho_{1}^{x}(d_{5}\rightarrow f_{3},d_{6}\rightarrow\tilde{f_{3}})}$}\tabularnewline
\cline{2-8}
 & $\ensuremath{\theta^{\pi}}$ & $\ensuremath{d_{2}+2d_{3}}$ & $\ensuremath{d_{2}+2d_{3}}$ & $\ensuremath{d_{2}+2d_{3}}$ & $\ensuremath{\frac{(d_{2}+6d_{3})}{4}}$ & $\ensuremath{\frac{1}{2}({d_{2}}+3{d_{3}})}$ & $0$\\
 & $\ensuremath{\theta^{k}}$ & $\ensuremath{d_{2}+2d_{3}}$ & $\ensuremath{{{\frac{1}{2}({d_2}+4{d_3})}}}$ & $2d_{3}$ & $\ensuremath{\frac{1}{2}(d_{2}+2d_{3})}$ & $\frac{1}{2}\left(d_{2}+2d_{3}\right)$ & $d_{2}+4d_{3}$\\
 & $\ensuremath{\theta^{\eta}}$ & $0$ & $0$ & $\ensuremath{0}$ & $\ensuremath{\frac{3}{4}({d_{2}}+2{d_{3}})}$ & $\ensuremath{\frac{3}{2}d_{3}}$ & $0$\tabularnewline
\cline{2-8}
 & $\ensuremath{X_{}^{\pi}}$ & $-1$ & $\ensuremath{0}$ & $\ensuremath{1}$ & $\ensuremath{\frac{1}{4}}$ & $\ensuremath{\frac{1}{2}}$ & $0$\\
 & $\ensuremath{X^{k}}$ & $-1$ & $\ensuremath{-\frac{1}{2}}$ & $\ensuremath{0}$ & $-\frac{1}{2}$ & $\ensuremath{\frac{1}{2}}$ & $1$\\
 & $\ensuremath{X^{\eta}}$ & $0$ & $0$ & $0$ & $\ensuremath{-\frac{1}{4}}$ & $0$ & $0$\tabularnewline
\bottomrule[1pt]\bottomrule[1pt]
\end{tabular}
\end{table}

\section{The Numerical results and discussions}\label{sec3}

There are fifteen LECs in the analytical expressions of the magnetic
moments up to $\mathcal{O}(p^3)$. In principle, they should be
determined by fitting the experiment data. So far, there is no
experiment data about the magnetic moments of the heavy baryons. Thus, as a second best scheme, we choose the Lattice QCD simulation data~\cite{Can:2013tna,Bahtiyar:2015sga,Can:2015exa,Bahtiyar:2016dom} as input. Before fitting the Lattice QCD results, we use the quark model or the heavy quark symmetry to reduce the number of unknown LECs.

The spin-$1\over 2$ and spin-$3\over 2$ sextet are degenerate states in the heavy quark limit. Their mass splitting is relatively small. The mass splitting between antitriplet and sextet are large. Thus, we do not take the antitriplet intermediate states into consideration when calculating the magnetic moments of sextet and vice versa. This issue will be discussed in detail in the following numerical results and Appendix~\ref{appn}. 

We will give the numerical results of the antitriplet charmed baryons to $\mathcal{O}(p^2)$, two Lattice QCD results are input. The numerical results of spin-$1\over 2$ sextet are given in two scenarios. In the first scenario, we use the Lattice QCD simulation data
\cite{Can:2013tna,Bahtiyar:2015sga,Can:2015exa,Bahtiyar:2016dom} and the quark
model to estimate these LECs. In the second scenario, we use the heavy quark symmetry  to reduce the independent LECs before fitting the Lattice QCD results.

\begin{table}
 \caption{The magnetic moments of the heavy baryons in $\bar{3}_f $ and the transition magnetic moments from $6_f$ to $\bar{3}_f $ (in units of $\mu_N$). }\label{3ft}
\begin{tabular}{c|cc|c|cc}
\toprule[1pt]\toprule[1pt] $\bar{3}_{f}$ & Quark model &
$\mathcal{O}(p^{1})$ & $6_{f}\rightarrow\bar{3}_{f}$ & Quark model &
$\mathcal{O}(p^{1})$\tabularnewline \midrule[1pt]
$\ensuremath{\mu_{\Lambda_{c}^{+}}}$ & $\mu_{c}$ &
$\frac{1}{2}\left(d_{2}+2d_{3}\right)$ &
$\mu_{\Sigma_{c}^{+}\rightarrow\Lambda_{c}^{+}\gamma}$ &
$-\frac{1}{\sqrt{3}}(\mu_{u}-\mu_{d})$ &
$4\sqrt{2}f_{2}$\tabularnewline $\ensuremath{\mu_{\Xi_{c}^{+}}}$ &
$\mu_{c}$ & $\frac{1}{2}\left(d_{2}+2d_{3}\right)$ &
$\ensuremath{\mu_{\Xi_{c}^{'+}\rightarrow\Xi_{c}^{+}\gamma}}$ &
$-\frac{1}{\sqrt{3}}(\mu_{u}-\mu_{s})$ &
$4\sqrt{2}f_{2}$\tabularnewline

$\ensuremath{\mu_{\Xi_{c}^{0}}}$ & $\mu_{c}$ &
$\ensuremath{d_{3}}$&
$\mu_{\Sigma_{c}^{*+}\rightarrow\Lambda_{c}^{+}\gamma}$ &
$\ensuremath{\frac{2}{\sqrt{6}}(\mu_{u}-\mu_{d})}$ &
$-\sqrt{\frac{2}{3}}\frac{f_{4}}{\sqrt{2}}$\tabularnewline
 & & & $\ensuremath{\mu_{\Xi_{c}^{*+}\rightarrow\Xi_{c}^{+}\gamma}}$ & $\frac{2}{\sqrt{6}}(\mu_{u}-\mu_{s})$ & $-\sqrt{\frac{2}{3}}\frac{f_{4}}{\sqrt{2}}$\tabularnewline
\bottomrule[1pt]\bottomrule[1pt]
\end{tabular}
\end{table}

\begin{table}
\caption{The magnetic moments of the heavy baryons in the $\bar 3_f$ representation order
by order (in units of $\mu_N$). Only the intermediate heavy baryons in the $\bar 3_f$ representation are included in
the chiral loops. }\label{m3}
\begin{center}
\begin{tabular}{c|ccl}
\toprule[1pt]\toprule[1pt] $\bar{3}_{f}$ & ~~$\mathcal{O}(p)$ ~~&
~~$\mathcal{O}(p^{2})$~~ &~~ Total \tabularnewline \midrule[1pt]
$\mu_{\Lambda_{c}^+}$ & 0.24 & 0 &
$0.24_{-0.02}^{+0.02}$ \tabularnewline
$\mu_{\Xi_{c}^{+}}^{\ddagger}$ & 0.24 & 0 &
$0.24_{-0.02}^{+0.02}$ \tabularnewline

$\mu_{\Xi_{c}^0}^{\ddagger}$ & 0.19 & 0 & $0.19_{-0.02}^{+0.02}$
\tabularnewline \bottomrule[1pt]\bottomrule[1pt]
\end{tabular}
\par\end{center}
\end{table}

\begin{table}
\scriptsize
 \caption{The(transition) magnetic moments of the heavy baryons in the sextet (in units of $\mu_N$).
 }\label{6ft}
\begin{tabular}{c|cc|c|cc|c|cc}
\toprule[1pt]\toprule[1pt] Spin-$\ensuremath{\frac{1}{2}}$ & Quark
model & $\mathcal{O}(p^{1})$ & Spin-$\ensuremath{\frac{3}{2}}$ &
Quark model & $\mathcal{O}(p^{1})$ &
$\ensuremath{\frac{3}{2}\rightarrow\frac{1}{2}}$ & Quark model &
$\mathcal{O}(p^{1})$\tabularnewline \midrule[1pt]
$\mu_{\Sigma_{c}^{++}}$ & $\frac{4}{3}\mu_{u}-\frac{1}{3}\mu_{c}$ &
$d_{5}+d_{6}$ & $\mu_{\Sigma_{c}^{*++}}$ & $2\mu_{u}+\mu_{c}$ &
$2(d_{8}+d_{9})$ &
$\ensuremath{\mu_{\Sigma_{c}^{*++}\rightarrow\Sigma_{c}^{++}}}$ &
$\frac{\sqrt{2}}{3}(2\mu_{\mu}-2\mu_{c})$ &
$\ensuremath{-\sqrt{\frac{2}{3}}(f_{3}+\tilde{f_{3}})}$\tabularnewline
$\mu_{\Sigma_{c}^{0}}$ &
$\ensuremath{\frac{4}{3}\mu_{d}-\frac{1}{3}\mu_{c}}$ &
$\ensuremath{d_{6}}$ & $\mu_{\Sigma_{c}^{*0}}$ & $2\mu_{d}+\mu_{c}$
& $2\ensuremath{d_{9}}$ &
$\mu_{\Sigma_{c}^{*0}\rightarrow\Sigma_{c}^{0}}$ &
$\ensuremath{\frac{\sqrt{2}}{3}(2\mu_{d}-2\mu_{c})}$ &
$\ensuremath{-\sqrt{\frac{2}{3}}\tilde{f_{3}}}$\tabularnewline

$\mu_{\Sigma_{c}^{+}}$ &
$\frac{2}{3}\mu_{u}+\frac{2}{3}\mu_{d}-\frac{1}{3}\mu_{c}$ &
$\frac{d_{5}}{2}+d_{6}$ & $\mu_{\Sigma_{c}^{*+}}$ &
$\mu_{u}+\mu_{d}+\mu_{c}$ & $2(\frac{d_{8}}{2}+d_{9})$ &
$\ensuremath{\mu_{\Sigma_{c}^{*+}\rightarrow\Sigma_{c}^{+}}}$ &
$\ensuremath{\frac{\sqrt{2}}{3}(\mu_{u}+\mu_{d}-2\mu_{c})}$ &
$\ensuremath{-\sqrt{\frac{2}{3}}(\frac{f_{3}}{2}+\tilde{f_{3}})}$\tabularnewline

$\mu_{\Xi_{c}^{'+}}$ &
$\ensuremath{\frac{2}{3}\mu_{u}+\frac{2}{3}\mu_{s}-\frac{1}{3}\mu_{c}}$
& $\frac{d_{5}}{2}+d_{6}$ & $\mu_{\Xi_{c}^{*+}}$ &
$\mu_{u}+\mu_{s}+\mu_{c}$ & $2(\frac{d_{8}}{2}+d_{9})$ &
$\ensuremath{\mu_{\Xi_{c}^{*'+}\rightarrow\Xi_{c}^{'+}}}$ &
$\ensuremath{\frac{\sqrt{2}}{3}(\mu_{u}+\mu_{s}-2\mu_{c})}$ &
$\ensuremath{-\sqrt{\frac{2}{3}}(\frac{f_{3}}{2}+\tilde{f_{3}})}$\tabularnewline

$\mu_{\Xi_{c}^{'0}}$ &
$\frac{2}{3}\mu_{d}+\frac{2}{3}\mu_{s}-\frac{1}{3}\mu_{c}$ & $d_{6}$
& $\mu_{\Xi_{c}^{*0}}$ & $\mu_{d}+\mu_{s}+\mu_{c}$ & $2d_{9}$ &
$\ensuremath{\mu_{\Xi_{c}^{*0}\rightarrow\Xi_{c}^{'0}}}$ &
$\frac{\sqrt{2}}{3}(\mu_{s}+\mu_{d}-2\mu_{c})$ &
$\ensuremath{-\sqrt{\frac{2}{3}}\tilde{f_{3}}}$\tabularnewline

$\mu_{\Omega_{c}^{0}}$ &
$\ensuremath{\frac{4}{3}\mu_{s}-\frac{1}{3}\mu_{c}}$ & $d_{6}$ &
$\mu_{\Omega_{c}^{*0}}$ & $2\mu_{s}+\mu_{c}$ & $2d_{9}$ &
$\ensuremath{\mu_{\Omega_{c}^{*0}\rightarrow\Omega_{c}^{0}}}$ &
$\frac{\sqrt{2}}{3}(2\mu_{s}-2\mu_{c})$ &
$-\sqrt{\frac{2}{3}}\tilde{f_{3}}$\tabularnewline
\bottomrule[1pt]\bottomrule[1pt]
\end{tabular}
\end{table}

\subsection{The charmed baryons in $\bar{3}_f$}

Before giving numerical results of antitriplet charmed baryons, it is heuristic to see the quark model results in Table~\ref{3ft}. All their magnetic moments are $\mu_c$. In the quark model, the light quarks do not contribute to the magnetic moments of $\bar{3}_f$ charmed baryons.

Within HBChPT, the analytical expressions up to $\mathcal{O}(p^2)$ contain two unknown coefficients, $d_2$ and $d_3$. The
$\mu_{\Xi^+_c}$ and the $\mu_{\Xi^0_c}$ from the Lattice QCD
simulations \cite{Bahtiyar:2016dom} are treated as input. The results are given in Table~\ref{m3}.  The errors come from the uncertainties of the Lattice QCD results. At the leading order, we separate the contribution of light quarks from the total magnetic moments,
\begin{equation}\label{lc}
\mu^{qq}_{\Lambda_c^+}=\mu^{qq}_{\Xi_c^+}=0.02 \mu_N,\quad \mu^{qq}_{\Xi_c^0}=-0.03 \mu_N,
\end{equation}
where the superscript $qq$ denotes the contribution from two light quarks. The calculation details are listed in Appendix \ref{apptraceless}. The contribution of light quarks is very small.
At $\mathcal{O}(p^2)$, the loop diagrams
with the intermediate $\bar 3_f$ states vanish because of $g_6=0$. Thus, at this order, the contribution from light degrees of freedom vanishes. Even if we take the sextet as the intermediate states in the loops, the contribution at this order is quite small as illustrated in Table~\ref{m3woall}. Thus, within HBChPT, we can also conclude that the heavy quark contribution dominates the magnetic moments of antitriplet charmed baryons.

\subsection{The charmed baryon in $6_f$: Scenario I}

We have calculated the analytical results for the magnetic moments of
the spin-$1\over 2$ sextet heavy baryons up to $\mathcal O(p^3)$. In addition to
$d_2$ and $d_3$, the analytical expressions contain the other eleven
parameters: $d_{4}$, $d_5$, $d_6$, $d_{8}$, $d_{9}$, $f_{2}$,
$f_{3}$, $\tilde{f}_{3}$ , {$f_{4}$}, $s_{2}$ and $s_{3}$. In this scenario, we use
the predictions from the quark model as the antitriplet magnetic
moments in the HBChPT and obtain the values of $d_5$, $d_6$,
$d_8$, $d_9$, $f_2$, $f_3$, $\tilde{f}_3$, and $f_4$. We obtain the
(transition) magnetic moments in the quark model and the leading-order (transition) magnetic moments, and list the analytical
expressions for the antitriplet and the sextet heavy baryons in
Table~\ref{3ft} and~\ref{6ft}, respectively. More details about quark model are
illustrated in Appendix~\ref{app3} and~\ref{app2}. In this work, we use the following the constituent quark masses in
the quark model,
\begin{eqnarray}
m_u=m_d=0.336~ \text{GeV}, ~ m_s=0.540~ \text{GeV},~ m_c=1.660~ \text{GeV}.
\end{eqnarray}
We use the
Lattice data $\mu_{\Sigma_c^{++}}$, $\mu_{\Xi^{'+}_c}$, and
$\mu_{\Omega_c^0}$ in Refs.
\cite{Can:2015exa,Bahtiyar:2015sga,Bahtiyar:2016dom,Can:2013tna} as input to
determine the other LECs, $d_4$, $s_2$, and $s_3$.

Both the
spin-$\frac{1}{2}$ and the spin-$\frac{3}{2}$ heavy baryons in the sextet are included as the
intermediate states in the loops. The numerical results are listed in Table
\ref{case1}. In calculation, we assume $10\%$ uncertainty for the quark mass.
This uncertainty together with the uncertainty of the Lattice data
leads to the errors in the numerical results. The values of the
parameters are listed in Table~\ref{LECS}. We obtain the
$\mu_{\Sigma_{c}^{+}}=0.26_{-0.07}^{+0.07} \mu_N$ ,
$\mu_{\Sigma_{c}^{0}}=-0.97_{-0.04}^{+0.05} \mu_N$, and
$\mu_{^{\Xi_{c}'^0}}=-0.84_{-0.03}^{+0.02} \mu_N$. Now the convergence of the chiral expansion works well. 

In Appendix~\ref{appn}, we include different intermediate states to investigate their contributions to the magnetic moments. By comparing the numerical results, we find that
the inclusion of the antitriplet heavy baryons as the intermediate
states will not change the final results of spin-$1\over 2$ sextet significantly. But it worsens the chiral convergence due to the large mass splitting between the antitriplet and sextet charmed baryons. Thus, we do not include the $\bar{3}_f$ intermediate states in the numerical analysis. More details are referred to Appendix~\ref{appn}.

\subsection{The charmed baryon in $6_f$: Scenario II}
In the heavy quark limit, the spin-$\frac{1}{2}$ and spin-$\frac{3}{2}$
sextet states are degenerate and we can relate some LECs to others with the heavy quark spin symmetry. In the heavy quark limit, the mass splitting is $\delta_2=0$  now. The two sextet states are combined into the following superfield~\cite{Cheng:1993kp,Cho:1992nt},
\begin{eqnarray}
&\psi^{\mu}=B_{6^{*}}^{\mu}-\sqrt{\frac{1}{3}}(\gamma^{\mu}+v^{\mu})\gamma_{5}B_{6},\nonumber \\
&\bar{\text{\ensuremath{\psi}}}{}_{\mu}=\bar{B}_{6^{*}}^{\mu}+\sqrt{\frac{1}{3}}\bar{B}_{6}\gamma_{5}(\gamma_{\mu}+v_{\mu}).
\end{eqnarray}
The $\mathcal{O}(p^{2})$
Lagrangians for the sextet electromagnetic interaction read~\cite{Cheng:1993kp,Cho:1992nt,Falk:1991nq,Banuls:1999br},

\begin{table}
\caption{The magnetic moments of the heavy baryons in the sextet order by
order (in units of $\mu_N$) in scenario I. The intermediate heavy baryons in the spin-$1\over2 $ and spin-$3\over2 $ sextet are included in the
chiral loops. The superscript $\ddagger$ indicates that the
corresponding Lattice data is treated as input. }\label{case1}
\begin{center}
\begin{tabular}{l|cccr}
\toprule[1pt]\toprule[1pt]
S-I & $\mathcal{O}(p)$ & $\mathcal{O}(p^{2})$ & $\mathcal{O}(p^{3})$ & Total\tabularnewline
 \midrule[1pt]
$\mu_{\Sigma_{c}^{++}}^{\ddagger}$ & $2.36$ & $-0.69$ & $-0.16$ & $1.50_{-0.19}^{+0.18}$\tabularnewline
$\mu_{\Sigma_{c}^{+}}$ & $0.61$ & $-0.26$ & $-0.09$ & $0.26_{-0.07}^{+0.07}$\tabularnewline
$\mu_{\Sigma_{c}^{0}}$ & $-1.13$ & $0.18$ & $-0.02$ & $-0.97_{-0.04}^{+0.05}$\tabularnewline
$\mu_{\Xi_{c}^{'+}}^{\ddagger}$ & $0.61$ & $-0.09$ & $-0.21$ & $0.32_{-0.12}^{+0.10}$\tabularnewline
$\mu_{\Xi_{c}^{'0}}$ & $-1.13$ & $0.35$ & $-0.06$ & $-0.84_{-0.03}^{+0.02}$\tabularnewline
$\mu_{\Omega_{c}^{0}}^{\ddagger}$ & $-1.13$ & $0.52$ & $-0.07$ & $-0.69_{-0.03}^{+0.03}$\tabularnewline
\bottomrule[1pt]\bottomrule[1pt]
\end{tabular}
\par\end{center}
\end{table}

\begin{table}
\caption{The fitted LECs in scenario I. }\label{LECS}
\begin{tabular}{cc|cc|cc}
\toprule[1pt]\toprule[1pt] LECs & Value & LECs & Value & LECs &
Value\tabularnewline \midrule[1pt] $d_{2}$ & $0.10$
& $d_{8}$ & $2.62$ & $f_{4}$ &
$-3.63$ \tabularnewline $d_{3}$ &
$0.19$ & $d_{9}$ & $-0.57$ & $s_{2}$
& $-0.19$\tabularnewline $d_{4}$ &
$3.45$& $f_{2}$ & $-0.27$ & $s_{3}$
& $-0.14$\tabularnewline $d_{5}$ &
$3.49$ & $f_{3}$ & $-3.00$ & &
\tabularnewline $d_{6}$ & $-1.13$ & $\tilde{f}_{3}$
& $1.30$ & & \tabularnewline
\bottomrule[1pt]\bottomrule[1pt]
\end{tabular}
\end{table}

\begin{eqnarray}
&&\mathcal{L}^{(2)}_{HQSS}=i\frac{g}{M_{N}}\bar{\psi}^{\mu}\hat{f}^{\mu\nu+}\psi_{\nu}+\frac{g_{t}}{M_{N}}\epsilon_{\mu\nu\alpha\beta}(\bar{\psi}^{\mu}\hat{f}^{\alpha\beta}v^{\nu}B_{\bar{3}}),\\
&&\mathcal{L}_{HQ}=\frac{g_{b}}{M_{N}}\bar{\psi}^{\lambda}\sigma_{\mu\nu}\psi_{\lambda}\text{Tr}(f^{\mu\nu+}),
\end{eqnarray}
where the $\hat{f}_{\mu\nu}^+=f_{\mu\nu}^+-\frac{1}{3}\text{Tr}(f_{\mu\nu}^+)$
is traceless. The $\mathcal{L}^{(2)}_{HQSS}$ is invariant under heavy quark spin transformation,  and represents the contribution
from the light quarks. The $\mathcal{L}_{HQ}$ violates the heavy quark spin symmetry and is related to the heavy quark contribution. 

By comparing these Lagrangians with those in Eqs. (\ref{wl5})
and (\ref{wl7}), we obtain,

\begin{eqnarray}
&d_{5}=-\frac{8}{3}g,\quad d_{8}=-2g,\quad f_{3}=4\sqrt{\frac{1}{3}}g,\nonumber\\
&d_9+ {1\over 3} d_8=- 4 g_b,\quad d_6 +{1\over 3} d_5= {8\over 3} g_b,\quad\tilde{f}_3+{1\over 3}{f_3} =- {16\over \sqrt{3}} g_b,\nonumber\\
 & f_{4}=8g_{t},\quad f_{2}=\frac{1}{\sqrt{3}}g_{t}.
\end{eqnarray}
The six LECs ($d_{5}$, $d_{6}$, $d_{8}$, $d_{9}$, $f_{3}$,
$\tilde{f}_{3}$) are then related to two independent LECs $g$ and $g_{b}$
in the heavy quark limit. And the $f_{2}$ and $f_{4}$ are related
to $g_{t}$. 
In this scenario, we consider the sextet heavy baryons as the intermediate states. Then there are four unknown LECs: $g$, $g_t$, $g_b$, and $s_3$. In Lattice QCD calculation~\cite{Can:2015exa,Bahtiyar:2015sga,Bahtiyar:2016dom,Can:2013tna},
the authors have given the contributions of the heavy quarks to magnetic moments:
\begin{equation}
\mu_{\Sigma^{++}}^{c}=-0.066\mu_{N},\quad
\mu_{\Xi_{c}^{'+}}^{c}=-0.059\mu_{N},\quad \mu_{\Omega_{c}^{0}}^{c}=-0.061\mu_{N}.
\end{equation}
We fit the average value $\mu_{6}^{c}=-0.06 \mu_N$ to obtain the $g_b$. We fit the remaining light quark contribution to determine $g_t$, $g_b$, and $s_3$. The numerical results are given up to
$\mathcal{O}(p^{3})$ in Tables~\ref{scenario2} and the corresponding values of LECs are listed in Tables~\ref{parameter2}.

\begin{table}
\caption{The magnetic moments of the charmed baryons in the sextet (in units of $\mu_N$) in scenario II. The intermediate heavy baryons in the spin-$1\over2 $ and spin-$3\over2 $ sextet are included in the chiral loops. The contributions from the light quarks to the magnetic moments are listed order by order. Those from the heavy quarks are listed in the fifth column. They combined to the total values listed in the last column. The superscript $\ddagger$ indicates that the
corresponding Lattice data is treated as input. }\label{scenario2}
\begin{center}
\begin{tabular}{c|ccccccc}
\toprule[1pt]\toprule[1pt]
S-II & $\mathcal{O}(p)$ & $\mathcal{O}(p^{2})$ & $\mathcal{O}(p^{3})$ & Heavy quark & Total\tabularnewline
\midrule[1pt]$\mu_{\Sigma_{c}^{++}}^{\ddagger}$ & $1.87$ & $-0.74$ & $0.42$& $-0.06$ & $1.50_{-0.20}^{+0.20}$ & \tabularnewline

$\mu_{\Sigma_{c}^{+}}$ & $0.47$ & $-0.26$ & $0.15$ & $-0.06$& $0.30_{-0.08}^{+0.09}$ & \tabularnewline
$\mu_{\Sigma_{c}^{0}}$ & $-0.94$ & $0.21$ & $-0.13$ & $-0.06$& $-0.91_{-0.22}^{+0.23}$ &\tabularnewline
$\mu_{^{\Xi_{c}'^{+}}}^{\ddagger}$ & $0.47$ & $-0.11$ & $0.01$ & $-0.06$& $0.31_{-0.15}^{+0.14}$ & \tabularnewline
$\mu_{^{\Xi_{c}'^0}}$ & $-0.94$ & $0.37$ & $-0.18$ & $-0.06$& $-0.80_{-0.11}^{+0.11}$ \tabularnewline
$\mu_{\varOmega_{c}^0}^{\ddagger}$ & $-0.94$ & $0.52$ & $-0.21$& $-0.06$ & $-0.69_{-0.03}^{+0.03}$ & \tabularnewline
\bottomrule[1pt]\bottomrule[1pt]
\end{tabular}
\par\end{center}
\end{table}

\begin{table}
\caption{The fitted LECs in scenario II. The $g$ and $g_b$ are related to the contributions of the light and heavy quarks, respectively. }\label{parameter2}
\begin{center}
\begin{tabular}{cc|cc|cc}
\toprule[1pt]\toprule[1pt]
\hline 
LECs & Value & LECs & Value & LECs & Value\tabularnewline
\midrule[1pt]
$g$ & $-1.05$ & $d_{5}$ & $2.81$ & $f_{3}$ & $-2.43$\tabularnewline
$d_{4}$ & $0.52$ & $d_{6}$ & $-0.98$ & $\tilde{f}_{3}$ & $0.97$\tabularnewline
$s_{3}$ & $-0.12$ & $d_{8}$ & $2.11$ & $g_{b}$ & $0.02$\tabularnewline
$s_{2}$ & $0.05$ & $d_{9}$ & $-0.63$ & & \tabularnewline
\bottomrule[1pt]\bottomrule[1pt]
\end{tabular}
\par\end{center}
\end{table}

\begin{table*}[ht]
\label{lattice}
 \caption{ The magnetic moments of the heavy baryons are given in different models. S-I and S-II denote the scenario I and scenario II (in units of $\mu_N$), respectively. The superscript $\ddagger$ denotes that the corresponding the Lattice data is treated as input.}\label{numerical result}
\begin{center}
 {
 \setlength{\extrarowheight}{5pt}
\begin{tabular*}{1.0\textwidth}{@{\extracolsep{\fill}}ccccccccccccccc}
 \hline\hline
 & S-I & S-II & {Lattice~\cite{Can:2015exa,Bahtiyar:2015sga,Can:2013tna,Bahtiyar:2016dom}}&~\cite{JuliaDiaz:2004vh}&~\cite{Faessler:2006ft} &~\cite{Sharma:2010vv} &~\cite{Barik:1984tq} &\cite{Bernotas:2012nz} &~\cite{Zhu:1997as} & \cite{Kumar:2005ei} & \cite{Patel:2007gx}
 &\cite{Yang:2018uoj} \\
 \hline \hline
 $\mu_{\Lambda_c^+}$ &\multicolumn{2}{c}{$0.24$} &- & 0.41 & 0.42 & 0.392 & 0.341 & 0.411 & - & 0.37 & 0.385 & -\\
 $\mu^{\ddagger}_{\Xi_c^+}$\cite{Bahtiyar:2016dom} &\multicolumn{2}{c}{$0.24$} & $0.235(25)$ & 0.39 & 0.41 & 0.40 & 0.341& 0.257 & - & 0.37 & - &-\\
 $\mu^{\ddagger}_{\Xi_c^0}$\cite{Bahtiyar:2016dom} &\multicolumn{2}{c}{$0.19$} & $0.192(17)$ & 0.39 & 0.39 & 0.28 & 0.341& 0.421 & - & 0.36 & - &-\\
 
 $\mu^{\ddagger}_{\Sigma^{++}_{c}}$ &$1.50$ & $1.50$ &1.499(202) & 3.07 & 1.76 & 2.20 & 2.44& 1.679 & 2.1(3) & 2.18 & 2.279 &$2.15(10)$\\
 $\mu_{\Sigma^{+}_{c}}$ & $0.26$&$0.30$ &- & 0.65 & 0.36 & 0.30 & 0.525& 0.318 & - & 0.63 & 0.501 & $0.46(3)$\\

 $\mu_{\Sigma^{0}_{c}}$& $-0.97$& $-0.91$ & -0.875(103) & -1.78 & -1.04 &-1.60 & -1.391 &- 1.043 & -1.6(2) & -1.17 & - 1.015 &$-1.24(5)$\\
 $\mu^{\ddagger}_{\Xi_c^{'+}}$\cite{Bahtiyar:2016dom}& $0.32$& $0.31$ & $0.315(141)$ & 1.13 & 0.47 & 0.76 & 0.796 & 0.591 & - & 0.76 & 0.711 &$0.60(2)$\\

 $\mu_{\Xi_c^{'0}}$\cite{Bahtiyar:2016dom} & $-0.84$& $-0.80$& $-0.599(71)$& -1.51 & -0.95 & -1.32 & - 1.12 &-0.914 & - & -0.93 & - 0.950 &$-1.05(4)$\\
 $\mu^{\ddagger}_{\Omega^{0}_{c}}$ & $-0.69$ & $-0.69$ &$-0.688(31)$ &- 0.90 & -0.85 &- 0.90 &- 0.85 & - 0.774 & - & -0.92 & - 0.960 & $-0.85(5)$ \\
 \hline\hline
 \end{tabular*}
 }
 \end{center}
 \end{table*}

\section{Summary} \label{sec4}

In summary, we have derived the analytical expressions of the
magnetic moments of spin-$1 \over 2$ singly charmed baryons up
to the next-to-next-to-leading order. We have performed the
calculation order by order in the framework of heavy baryon chiral
perturbation theory. There are several relations between the magnetic moments of the charmed baryons up to $\mathcal{O}(p^2)$. Most of them are not valid any more at $\mathcal O(p^3)$.
The number of LECs involved is larger than that of the magnetic moments to be calculated. We have used two scenarios to reduce and estimate these LECs.

We have obtained the numerical values of the magnetic moments
up to $\mathcal{O}(p^2)$ for the $\bar 3_f$ charmed baryon.  The light quarks  have little contribution to the magnetic moment. We have given the numerical results for the spin-$1\over 2$ sextet up to $\mathcal{O}(p^3)$ in two scenarios. In the first scenario, the LECs were estimated using the Lattice QCD data and quark model due to the lack of experiment
data. The convergence of the chiral expansion works well if we only consider
the sextet as the intermediate states in the loops. The inclusion of
the intermediate antitriplet charmed baryons worsens the convergence
and does not change the numerical results significantly. In the second scenario, the heavy quark symmetry was used to reduce the number of the independent LECs. 
The magnetic moments were decomposed into the heavy and light parts, respectively. With the numerical results of the Lattice QCD simulation as input, we have obtained the values of the LECs and the numerical results.

We have listed the numerical results in the above two scenarios in Table~\ref{numerical result}. The numerical results are similar to each other. The predicted values of $\mu_{\Sigma^0_c}$ and $\mu_{\Xi^{'0}_c}$ are consistent with those of the Lattice QCD simulation results. In this Table, we have also compared our numerical
results with those results in the Lattice QCD
\cite{Can:2015exa,Bahtiyar:2015sga,Can:2013tna,Bahtiyar:2016dom}, the
relativistic quark model~\cite{JuliaDiaz:2004vh}, the relativistic
three-quark model~\cite{Faessler:2006ft}, the chiral constituent
quark model ($\chi$ CQM)~\cite{Sharma:2010vv}, an independent-quark
model based on Dirac equation with power-law potential
\cite{Barik:1984tq}, the bag model~\cite{Bernotas:2012nz}, the QCD
sum rule~\cite{Zhu:1997as}, the effective mass and screened charge
scenario in the Ref.~\cite{Kumar:2005ei}, the hyper central model
\cite{Patel:2007gx}, and the mean-field approach
\cite{Yang:2018uoj}. 

It's very interesting to note that the results
from various models are roughly consistent with ours. The numerical
results of the heavy baryon magnetic moments from the Lattice
simulations are generally smaller than the quark model predictions.
Due to the lack of the experimental data, we use several Lattice
data as input to extract the low-energy constants, which renders
some of our results are also smaller than the quark model estimates.
With the analytical expressions derived in this work, we may further
improve and update the numerical analysis in the future if the
magnetic moments of several heavy baryons are measured
experimentally or more accurate Lattice QCD simulations become
available.

\section*{Acknowledgements}

The authors are grateful to X. L. Chen, W. Z. Deng and Meng-Lin Du
for useful discussions. This project is supported by the National
Natural Science Foundation of China under Grants No.11575008 and No.
11621131001 and the National Key Basic Research Program of
China(2015CB856700). This work is also supported by the Fundamental
Research Funds for the Central Universities of Lanzhou University
under Grants No. 223000-62637.

\begin{appendix}

\section{The leading-order (transition) magnetic moments}\label{app3}
With HBChPT, the leading-order magnetic moments of the
spin-$1\over 2$ heavy baryons are given in Eqs. (\ref {op13bar}) and
(\ref{op16}). The transition magnetic moments for the spin-$1\over
2$ sextet to the antitriplet heavy baryons are listed in Table
\ref{3ft}. For the spin-$\frac{3}{2}$ heavy baryons, the matrix
element of electromagnetic current is~\cite{Li:2016ezv},
\begin{eqnarray}
&\langle B^{*\rho}(p^{\prime})|\mathcal{J}_{\mu}|B^{*\sigma}(p)\rangle =\bar{u}^{\rho}(p^{\prime})\mathcal{O}_{\rho\mu\sigma}(p^{\prime},p)u^{\sigma}(p),\\
 &\mathcal{O}_{\rho\mu\sigma}(p^{\prime},p)=-g_{\rho\sigma}\left[v_{\mu}F_{1}(q^{2})+\frac{[S_{\mu},S_{\alpha}]}{M_{6^*}}q^{\alpha}F_{2}(q^{2})\right]+\frac{q_{\rho}q_{\sigma}}{(2M_{6^*})^{2}}\left[v_{\mu}F_{3}(q^{2})+\frac{[S_{\mu},S_{\alpha}]}{M_{6^*}}q^{\alpha}F_{4}(q^{2})\right],
\end{eqnarray}
where the transferred momentum $q=p^{\prime}-p$. $F_{1-4}$ are the functions of $q^2$. The magnetic-dipole (M1) form factor and the
magnetic moment are
\begin{eqnarray}
G_{M1}(q^{2})=(1+\frac{4}{5}\tau)F_{2}-\frac{2}{5}\tau(1+\tau)F_{4},\quad \mu_{B^*}=\frac{e}{2M_{6^*}}G_{M1}(0),
\end{eqnarray}
where $\tau=-\frac{q^{2}}{4M^{2}_{6^*}}$. The $\mu_{B^*}$ is the magnetic moments of the spin-$3\over 2$ heavy baryons and it can be
derived from the Lagrangians in Eq. (\ref{wl5}).

For the radiative decay of a spin-$3\over 2$ heavy baryon into a
spin-$1 \over 2$ one, the magnetic moment reads~\cite{Li:2017pxa,Jones:1972ky},
 \begin{eqnarray}
& \langle B^{ *\nu}_{6 }(p')| \mathcal{J}^{\mu}|B(p) \rangle=e\bar u^{\nu}(p')\left[\frac{G_1}{M_B}(q^{\nu} S^{\mu}-q\cdot Sg^{\nu\mu} )+\frac{G_2}{4M^2_B}(q^{\nu}v^{\mu}-q\cdot v g^{\nu \mu})q\cdot S\right]u(p)\nonumber\\
 &G_{M1}=\frac{2}{3}G_1-\frac{\delta}{6M_{6^*}}G_1+\frac{\delta}{12M_{B}}G_2\nonumber\\
 &\mu({B^*}\rightarrow {B\gamma})=-\sqrt{3\over 2}G_{M1}(q^2=0)\frac{e}{2M_B}.
\end{eqnarray}
Using this equation, we can derive the $\mu({B^*}\rightarrow {B\gamma})$
from the Lagrangians in Eq. (\ref{wl7}).

\section{Quark model}\label{app2}

The electromagnetic current at the quark level is
 \begin{eqnarray}
\mathcal{J}^{\mu}=\bar q Q_l \gamma_{\mu} q+\frac{2}{3}\bar
c\gamma_\mu c,
\end{eqnarray}
where $q=(u,d,s)^T$ is the light quark field and $c$ is the charm
quark field. In the quark model, the wave functions and the
corresponding magnetic moments of the heavy baryons in different
flavor representations read,

spin-${1\over 2}$ $\bar{3}_f$:
 \begin{eqnarray}
 \label{n1}
&~
|\frac{1}{2},\frac{1}{2}\rangle=\frac{1}{\sqrt{2}}(q_{1}q_{2}-q_{2}q_{1})Q\otimes\frac{1}{\sqrt{2}}(\uparrow\downarrow-\downarrow\uparrow)\uparrow,\qquad
\mu_{\bar 3}=2(\mu_{q_{1}}-\mu_{q_{2}})\nonumber
\end{eqnarray}
 spin-${1\over 2}$ $6_f$:
\begin{eqnarray}
 |\frac{1}{2},\frac{1}{2}\rangle&=&qqQ\otimes(-\frac{1}{\sqrt{3}}\frac{1}{\sqrt{2}}(\uparrow\downarrow+\downarrow\uparrow)\uparrow+\sqrt{\frac{2}{3}}\uparrow\uparrow\downarrow)
, ~~~~~~~~~~~~~~~~~~~~~~~\mu_6=\frac{4}{3}\mu_{q}-\frac{1}{3}\mu_{c}\nonumber \\
|\frac{1}{2},\frac{1}{2}\rangle&=&\frac{1}{\sqrt{2}}(q_{1}q_{2}+q_{2}q_{1})Q\otimes(-\frac{1}{\sqrt{3}}\frac{1}{\sqrt{2}}(\uparrow\downarrow+\downarrow\uparrow)\uparrow+\sqrt{\frac{2}{3}}\uparrow\uparrow\downarrow),~~~~\mu_6=\frac{1}{3}(2u_{q_{1}}+2\mu_{q_{2}}-\mu_{c})\nonumber
\end{eqnarray}
spin-${3\over 2}$ $6_f$:
\begin{eqnarray}
 |\frac{3}{2},\frac{3}{2}\rangle&=&qqQ\otimes\uparrow\uparrow\uparrow ,~ ~~~~~~~~~~~~~~~~~~~~~\mu_{6^*}=2\mu_{q}+\mu_{c}\nonumber \\
|\frac{3}{2},\frac{3}{2}\rangle&=&\frac{1}{\sqrt{2}}(q_{1}q_{2}+q_{2}q_{1})Q\otimes\uparrow\uparrow\uparrow,
~ ~~~\mu_{6^*}=\mu_{q_{1}}+\mu_{q_{2}}+\mu_{c}\nonumber
\end{eqnarray}
where $q_{i}$ and $Q$ are the light and heavy quarks in the heavy
baryon as illustrated in Fig.~\ref{w2}, respectively. The
$\uparrow(\downarrow)$ represents the direction of the third
component of the quark spin. $\mu_{q(c)}=\frac{e_{q(c)}}{2m_{q(c)}}$
is the quark magnetic moment. The transition magnetic moments in the
quark model are,
 \begin{eqnarray}
&6_f ~\text{with~spin}~ \frac{1}{2}\rightarrow \bar 3_f :
&\mu_{6\rightarrow\bar 3}=-\frac{1}{\sqrt{3}}(\mu_{q_1}-\mu_{q_2})\\
&6_f ~\text{with~spin}~ \frac{3}{2}\rightarrow \bar 3_f :
& \mu_{6^*\rightarrow\bar 3}=\frac{2}{\sqrt{6}}(\mu_{q_1}-\mu_{q_2})\\
&6_f ~\text{with~spin}~ \frac{3}{2}\rightarrow6_f ~\text{with~spin}~
\frac{1}{2}: & \mu_{6^*\rightarrow 6}=\frac{\sqrt
2}{{3}}(\mu_{q_1}+\mu_{q_2}-2\mu_Q)
\end{eqnarray}

\section{The effect of different intermidate states}\label{appn}

For the the magnetic moments of the
heavy baryons in the antitriplet, the light quarks do not contribute since the total light-quark spin
$S_{l}=0$. Their magnetic moments are $\mu_c$ as illustrated in
Table \ref{3ft}. Within HBChPT, the contribution to the magnetic
moment from the light quark at the leading order comes from the
$d_{2}$ term. Thus, if we treat the predictions in the quark model
as the leading-order magnetic moments, we get $d_2=0$. However, we
notice that the chiral expansion suffers from bad convergence with
the above treatment because of two reasons. Firstly, $\mu_c$ is
proportional to the $1/m_c$, which is small and of the same order
with $1/M_B$. In the HBChPT Lagrangian, we have dropped off the
$1/M_B$ terms. Thus, it is not consistent to fit the $d_2$ and $d_3$
using the quark model results. Secondly, the chiral corrections for
the $\mathcal{O}(p^2)$ magnetic moments are also quite small as
illustrated in the Tables \ref{m3}  and \ref{m3woall}. At this order, the loop diagrams
with the intermediate $\bar 3_f$ states should give the major chiral
correction. However, these diagrams vanish due to $g_6=0$. Moreover,
the opposite contributions from the spin-$1\over 2$ and the
spin-$3\over 2$ sextet heavy baryons almost cancel out. The above reasons make the convergence of the chiral
expansion quite uncontrollable.

\begin{table}
\caption{The magnetic moments of the heavy baryons in the $\bar 3_f$ representation order
by order (in units of $\mu_N$). The intermediate heavy baryons in the spin-$\frac{1}{2}$ and
spin-$\frac{3}{2}$ sextet are included in
the chiral loops. }\label{m3woall}
\begin{center}
\begin{tabular}{c|ccl}
\toprule[1pt]\toprule[1pt] $\bar{3}_{f}$ & ~~$\mathcal{O}(p)$ ~~&
~~$\mathcal{O}(p^{2})$~~ &~~ total \tabularnewline \midrule[1pt]
$\mu_{\Lambda_{c}^+}$ & 0.19 & 0.02 & $0.21$
\tabularnewline

$\mu_{\Xi_{c}^{+}}^{\ddagger}$ & 0.19 & 0.05 &
$0.24$ \tabularnewline

$\mu_{\Xi_{c}^0}^{\ddagger}$ & 0.25 & -0.06 & $0.19$
\tabularnewline \bottomrule[1pt]\bottomrule[1pt]
\end{tabular}
\par\end{center}
\end{table}


 In order to investigate the effect of different intermediate states on the final results and chiral convergence, we give the magnetic moments of the spin-$1\over 2$ sextet charmed baryons with another method  in scenario I. Both the antitriplet and the sextet charmed baryons are
included as the intermediate states in the loops. The numerical
results are listed in Table~\ref{case2}. Comparing Tables~\ref{case1}
with~\ref{case2}, we notice that the addition of the $\bar 3_f$ intermediate heavy baryons worsens the
convergence of the chiral expansion.
\begin{table}
\caption{The magnetic moments of the charmed baryons in the spin-$1\over 2$ sextet order by order (in units of $\mu_N$). The intermediate charmed baryons in the antitriplet and the sextet are all included in the chiral loops. }\label{case2}
\begin{center}
\begin{tabular}{l|cccr}
\toprule[1pt]\toprule[1pt]

S-I & ~~$\mathcal{O}(p)$~~ &~~ $\mathcal{O}(p^{2})$~~ &~~ $\mathcal{O}(p^{3})$ ~~ & ~~Total\tabularnewline
 \midrule[1pt]

$\mu_{\Sigma_{c}^{++}}^{\ddagger}$ & $2.36$ & $-1.01$ & $0.15$ & $1.50$\tabularnewline
$\mu_{\Sigma_{c}^{+}}$ & $0.61$ & $-0.50$ & $0.01$ & $0.12$\tabularnewline
$\mu_{\Sigma_{c}^{0}}$ & $-1.13$ & $0.01$ & $-0.14$ & $-1.27$\tabularnewline
$\mu_{\Xi_{c}^{'+}}^{\ddagger}$ & $0.61$ & $-0.004$ & $-0.29$ & $0.32$\tabularnewline
$\mu_{\Xi_{c}^{'0}}$ & $-1.13$ & $0.50$ & $-0.32$ & $-0.95$\tabularnewline
$\mu_{\Omega_{c}^{0}}^{\ddagger}$ & $-1.13$ & $1.00$ & $-0.56$ & $-0.69$\tabularnewline
 \bottomrule[1pt]\bottomrule[1pt]
\end{tabular}
\par\end{center}
\end{table}

\section{The contributions of the light and heavy quarks}\label{apptraceless}

The magnetic moments  of the charmed baryons are composed of the contributions from the light and charm quarks. 
We take the calculation of Eq. (32) as an example to illustrate the decomposition of these two contributions. At the leading order, the magnetic moments of the charmed baryons in the antitriplet representation arise from the ${\mathcal {L}}^{(2)}_{33}$ in  Eq. (\ref{wl5}).
We rewrite the Lagrangian as follows, 
\begin{eqnarray}
{\mathcal {L}}^{(2)}_{33}=-\frac{i\hat{d}_2}{8M_N}\text{Tr}(\bar B_{\bar{3}}[S^{\mu},S^{\nu}]\hat{f}_{\mu\nu}^{+}B_{\bar{3}})-\frac{i\hat{d}_3}{8M_N}\text{Tr}(\bar B_{\bar{3}}[S^{\mu},S^{\nu}]B_{\bar{3}})\text{Tr}(f_{\mu\nu}^{+}),
\end{eqnarray} 
where $\hat f_{\mu\nu}^+={f}_{\mu\nu}^+-\frac{1}{3}\text{Tr}(f_{\mu\nu}^+)$ is traceless. The  $\hat f_{\mu\nu}^+$ is related to the traceless charge matrix of the light quarks $Q_l=\text{diag}({{2\over 3}, -{1\over 3},-{1\over 3}})$. The $\text{Tr}(f_{\mu\nu}^+)$  is related to the charge matrix of the heavy quark $Q_c=\text{diag}({{1\over 3}, {1\over 3},{1\over 3}})$. Thus, the $\hat{d}_2$ and $\hat{d}_3$ term denote the light and heavy quarks' contributions, respectively. Combining the equation with  Eq. (\ref{wl5}), we obtain the relation between $\hat d _{2,3}$ and $d_{2,3}$,
\begin{eqnarray}
d_2=\hat{d}_2,~~~d_3={\hat d}_3-\frac{1}{3}\hat d_2.
\end{eqnarray}
Then, the analytical expressions of the leading-order magnetic moments in Eq. (\ref {op13bar})  can be expressed as, 
\begin{eqnarray}
\mu^{(1)}_{\Lambda_{c}^{+}}=\frac{1}{2}\left(\frac{1}{3}\hat d_{2}+2 \hat d_{3}\right), &
\mu^{(1)}_{\Xi_{c}^0}={\hat d}_3-\frac{1}{3}\hat d_2,
&\mu^{(1)}_{\Xi_{c}^{+}}=\frac{1}{2}\left(\frac{1}{3}\hat d_{2}+2 \hat d_{3}\right).\label{op13bartraceless}
\end{eqnarray}
Using the values of the $d_2$ and $d_3$ in Table VIII, we obtain $\hat d_2=0.10$ and $\hat d_3=0.22$. The contribution from the light quarks to the total magnetic moments are

\begin{equation}
\mu^{qq}_{\Lambda_c^+}=\mu^{qq}_{\Xi_c^+}=\frac{1}{6}\hat{d}_2 =0.02 \mu_N,\quad \mu^{qq}_{\Xi_c^0}=-\frac{1}{3}\hat d_2=-0.03 \mu_N.
\end{equation}

\section{Loop integrals}\label{app1}

The integral definitions in Eqs. (\ref{32}), (\ref{wl11}), and
(\ref{wl12}) are the same as those in Ref.~\cite{Li:2016ezv}.
 \begin{eqnarray}
i\int\frac{d^{d}l\lambda^{4-d}}{(2\pi)^{d}}\frac{[1,l_{\alpha},l_{\alpha}l_{\beta}]}{(l^{2}-m^{2}+i\epsilon)(w+v\cdot
l+i\epsilon)}=[J_{0}(w),v_{\alpha}J_{1}(w),g_{\alpha\beta}J_{2}(w)+v_{\alpha}v_{\beta}J_{3}(w)],
\end{eqnarray}

\begin{eqnarray}
i\int\frac{d^{d}l\lambda^{4-d}}{(2\pi)^{d}}\frac{[1,l_{\alpha},l_{\alpha}l_{\beta}]}{(l^{2}-m^{2}+i\epsilon)(w+v\cdot
l+i\epsilon)^{2}}=-[\frac{\partial J_{0}(w)}{\partial
w},v_{\alpha}\frac{\partial J_{1}(w)}{\partial
w},g_{\alpha\beta}J'_2(w)+v_{\alpha}v_{\beta}\frac{\partial
J_{3}(w)}{\partial w}], ~~
\end{eqnarray}

\begin{eqnarray}
i\int\frac{d^{d}l\lambda^{4-d}}{(2\pi)^{d}}\frac{[1,l_{\alpha},l_{\alpha}l_{\beta}]}{(l^{2}-m^{2}+i\epsilon)(v\cdot
l+i\epsilon)(w+v\cdot
l+i\epsilon)}=[\Gamma_{0}(w),v_{\alpha}\Gamma_{1}(w),g_{\alpha\beta}\Gamma_{2}(w)+v_{\alpha}v_{\beta}\Gamma_{3}(w)],~~
\end{eqnarray}

\begin{eqnarray}
i\int\frac{d^{d}l\lambda^{4-d}}{(2\pi)^{d}}\frac{[1,l_{\alpha},l_{\alpha}l_{\beta},l_{\nu}l_{\alpha}l_{\beta}]}{(l^{2}-m^{2}+i\epsilon)((l\text{+q})^{2}-m^{2}+i\epsilon)(w+v\cdot
l+i\epsilon)}=[L_{0}(w),L_{\alpha},L_{\alpha\beta},L_{\nu\alpha\beta}], \nonumber
~v\cdot q=0,
\end{eqnarray}

\begin{eqnarray}
L_{\alpha\beta}=n_{1}^{\textrm{II}}g_{\alpha\beta}+n_{2}^{\textrm{II}}q_{\alpha}q_{\beta}+n_{3}^{\textrm{II}}v_{\alpha}v_{\beta}+n_{4}^{\textrm{II}}v_{\alpha}q_{\beta}+n_{5}^{\textrm{II}}q_{\alpha}v_{\beta},
\end{eqnarray}

\begin{eqnarray}
n_{1}^{\textrm{II}}(-\delta,m)=\begin{cases}
\frac{m}{16\pi} & \delta=0\\
\frac{\text{-\ensuremath{\delta}}\left(\ln\left(\frac{m^{2}}{\lambda^{2}}\right)-1\right)+2\sqrt{m^{2}-\delta^{2}}\cos^{-1}\left(\frac{\delta}{m}\right)}{16\pi^{2}}
& \delta<m,
\end{cases}
\end{eqnarray}

\begin{eqnarray}
\frac{4}{d-1}n_{1}^{\textrm{II}}(-\delta,m)=\begin{cases}
\frac{-6\sqrt{\delta^{2}-m^{2}}\cosh^{-1}\left(\frac{\delta}{m}\right)-\delta\left(3\ln\left(\frac{m^{2}}{\lambda^{2}}\right)-5\right)}{36\pi^{2}} & \delta>m\\
\frac{5\delta+6\sqrt{m^{2}-\delta^{2}}\cos^{-1}\left(\frac{\delta}{m}\right)-3\delta\ln\left(\frac{m^{2}}{\lambda^{2}}\right)}{36\pi^{2}}
& \delta<m,
\end{cases}
\end{eqnarray}

\begin{eqnarray}
\frac{1-d}{4}J'(-\delta)=\begin{cases}
-\frac{3m^{2}\ln\left(\frac{m^{2}}{\lambda^{2}}\right)+2m^{2}}{64\pi^{2}} & \delta=0\\
-\frac{3\left(m^{2}-2\delta^{2}\right)\ln\left(\frac{m^{2}}{\lambda^{2}}\right)+2\left(\delta^{2}+m^{2}\right)+12\delta\sqrt{m^{2}-\delta^{2}}\cos^{-1}\left(\frac{\delta}{m}\right)}{64\pi^{2}} & \delta<m\\
\frac{-3\left(m^{2}-2\delta^{2}\right)\ln\left(\frac{m^{2}}{\lambda^{2}}\right)-2\left(\delta^{2}+m^{2}\right)+12\delta\sqrt{\delta^{2}-m^{2}}\cosh^{-1}\left(\frac{\delta}{m}\right)}{64\pi^{2}}
& \delta>m,
\end{cases}
\end{eqnarray}

\begin{eqnarray}
\frac{3-d}{4}J'_2(-\delta)=\begin{cases}
-\frac{-2\delta^{2}+\left(m^{2}-2\delta^{2}\right)\ln\left(\frac{m^{2}}{\lambda^{2}}\right)+4\delta\sqrt{m^{2}-\delta^{2}}\cos^{-1}\left(\frac{\delta}{m}\right)+2m^{2}}{64\pi^{2}} & \delta<m\\
\frac{2\delta^{2}-\left(m^{2}-2\delta^{2}\right)\ln\left(\frac{m^{2}}{\lambda^{2}}\right)+4\delta\sqrt{\delta^{2}-m^{2}}\cosh^{-1}\left(\frac{\delta}{m}\right)-2m^{2}}{64\pi^{2}}
& \delta>m,
\end{cases}
\end{eqnarray}
\begin{eqnarray}
(d-2)J'_2(-\delta)=\begin{cases}
\frac{\left(m^{2}-2\delta^{2}\right)\ln\left(\frac{m^{2}}{\lambda^{2}}\right)+4\delta\sqrt{m^{2}-\delta^{2}}\cos^{-1}\left(\frac{\delta}{m}\right)+m^{2}}{8\pi^{2}} & \delta<m\\
\frac{\left(m^{2}-2\delta^{2}\right)\ln\left(\frac{m^{2}}{\lambda^{2}}\right)-4\delta\sqrt{\delta^{2}-m^{2}}\cosh^{-1}\left(\frac{\delta}{m}\right)+m^{2}}{8\pi^{2}}
& \delta>m,
\end{cases}
\end{eqnarray}

\begin{eqnarray}
\left(\frac{4(5-d)}{(d-1)^{2}}+\frac{8}{1-d}\right)J'_2(-\delta)=\begin{cases}
\frac{-15\left(m^{2}-2\delta^{2}\right)\ln\left(\frac{m^{2}}{\lambda^{2}}\right)+2\left(m^{2}-17\delta^{2}\right)-60\delta\sqrt{m^{2}-\delta^{2}}\cos^{-1}\left(\frac{\delta}{m}\right)}{108\pi^{2}} & \delta<m\\
\frac{-15\left(m^{2}-2\delta^{2}\right)\ln\left(\frac{m^{2}}{\lambda^{2}}\right)+2\left(m^{2}-17\delta^{2}\right)+60\delta\sqrt{\delta^{2}-m^{2}}\cosh^{-1}\left(\frac{\delta}{m}\right)}{108\pi^{2}}
& \delta>m,
\end{cases}~~~
\end{eqnarray}

\begin{eqnarray}
\frac{3-d}{d-1}\frac{J_{2}(-\delta_{1})-J_{2}(-\delta_{2})}{\delta_{1}-\delta_{2}}=\begin{cases}
\frac{\left(9\delta_{2}m^{2}-6\delta_2^{3}\right)\ln\left(\frac{m^{2}}{\lambda^{2}}\right)+2\left(\delta_{2}^{3}+3\pi\left(m^{2}\right)^{3/2}\right)-12\left(m^{2}-\delta_{2}^{2}\right)^{3/2}\cos^{-1}\left(\frac{\delta_{2}}{m}\right)}{432\pi^{2}\delta_2} & \delta_{1}=0,\delta_{2}<m\\
\\
\frac{2\text{\ensuremath{\delta_{1}}}^{3}-2\text{\ensuremath{\delta_{2}}}^{3}-6\text{\ensuremath{\delta_{1}}}^{3}\ln\left(\frac{m^{2}}{\lambda^{2}}\right)-12\left(m^{2}-\text{\ensuremath{\delta_{1}}}^{2}\right)^{3/2}\cos^{-1}\left(\frac{\text{\ensuremath{\delta_{1}}}}{m}\right)+9\text{\ensuremath{\delta_{1}}}m^{2}\ln\left(\frac{m^{2}}{\lambda^{2}}\right)}{432\pi^{2}(\text{\ensuremath{\delta_1}}-\text{\ensuremath{\delta_2}})}\\
~~~+\frac{6\text{\ensuremath{\delta_{2}}}^{3}\ln\left(\frac{m^{2}}{\lambda^{2}}\right)+12\left(m^{2}-\text{\ensuremath{\delta_{2}}}^{2}\right)^{3/2}\cos^{-1}\left(\frac{\text{\ensuremath{\delta_{2}}}}{m}\right)-9\text{\ensuremath{\delta_{2}}}m^{2}\ln\left(\frac{m^{2}}{\lambda^{2}}\right)}{432\pi^{2}(\text{\ensuremath{\delta_1}}-\text{\ensuremath{\delta_2}})} & \delta_{1},\delta_{2}<m\\
\\
\frac{2\text{\ensuremath{\delta_{1}}}^{3}-2\text{\ensuremath{\delta_{2}}}^{3}-6\text{\ensuremath{\delta_{1}}}^{3}\ln\left(\frac{m^{2}}{\lambda^{2}}\right)-12\left(m^{2}-\text{\ensuremath{\delta_{1}}}^{2}\right)^{3/2}\cos^{-1}\left(\frac{\text{\ensuremath{\delta_{1}}}}{m}\right)+9\text{\ensuremath{\delta_{1}}}m^{2}\ln\left(\frac{m^{2}}{\lambda^{2}}\right)}{432\pi^{2}(\text{\ensuremath{\delta_1}}-\text{\ensuremath{\delta_2}})}\\
~~~+\frac{6\text{\ensuremath{\delta_{2}}}^{3}\ln\left(\frac{m^{2}}{\lambda^{2}}\right)+12\left(\text{\ensuremath{\delta_{2}}}^{2}-m^{2}\right)^{3/2}\cosh^{-1}\left(\frac{\text{\ensuremath{\delta_{2}}}}{m}\right)-9\text{\ensuremath{\delta_{2}}}m^{2}\ln\left(\frac{m^{2}}{\lambda^{2}}\right)}{432\pi^{2}(\text{\ensuremath{\delta_1}}-\text{\ensuremath{\delta_2}})}
& \delta_{1}<m,\delta_{2}>m,
\end{cases}
\end {eqnarray}
\begin{eqnarray}
 \Gamma_2(w)=\frac{J_2(0)-J_2(w)}{w}.\nonumber
 \end{eqnarray}

\end{appendix}

\end{document}